\documentclass[number,sort&compress,5p]{elsarticle}
\pdfoutput=1
\usepackage{amsmath,amssymb,enumitem,accents}
\usepackage{graphicx}
\usepackage{aas_macros}
\usepackage{float}
\def\beq{\begin{equation}}   \def\eeq{\end{equation}}
\def\noi{\noindent}

\begin{document}

\title{Extraction of black hole coalescence waveforms from noisy data}

\author[PI]{Martin A. Green}
\ead{mgreen@perimeterinstitute.ca}
\author[PI,UW]{J. W. Moffat}
\ead{jmoffat@perimeterinstitute.ca}
\address[PI]{Perimeter Institute for Theoretical Physics,
  Waterloo ON N2L 2Y5, Canada}

\address[UW]{Department of Physics and Astronomy, University of
  Waterloo, Waterloo ON N2L 3G1, Canada\\[8pt]\protect{\textrm
    \copyright~2018. This manuscript version is made available under
    the CC-BY-NC-ND 4.0 license\\
    http://creativecommons.org/licenses/by-nc-nd/4.0/}}

\begin{abstract}
  We describe an independent analysis of LIGO data for black hole
  coalescence events.  Gravitational wave strain waveforms are
  extracted directly from the data using a filtering method that
  exploits the observed or expected time-dependent frequency content.
  Statistical analysis of residual noise, after filtering out spectral
  peaks (and considering finite bandwidth), shows no evidence of
  non-Gaussian behaviour.  There is also no evidence of anomalous
  causal correlation between noise signals at the Hanford and
  Livingston sites.  The extracted waveforms are consistent with black
  hole coalescence template waveforms provided by LIGO.  Simulated
  events, with known signals injected into real noise, are used to
  determine uncertainties due to residual noise and demonstrate that
  our results are unbiased.  Conceptual and numerical differences
  between our RMS signal-to-noise ratios (SNRs) and the published
  matched-filter detection SNRs are discussed.\\[4pt]
  Declaration of interests: None\\[4pt]

\end{abstract}

\maketitle

\section{Introduction}\label{S:Intro}

The LIGO Scientific Collaboration and Virgo Collaboration (LIGO) have
reported six events in which spatial strain measurements are
consistent with gravitational waves produced by the inspiral and
coalescence of binary black holes~\cite{PhysRevLett.116.061102,
  PhysRevD.93.122003, PhysRevLett.116.241103, PhysRevX.6.041015,
  PhysRevLett.118.221101,PhysRevLett.119.141101,2017ApJ...851L..35A}.
In each case, similar signals were detected at the Hanford, Washington
(H) and Livingston, Louisiana (L) sites, with a time offset less than
the 10~ms inter-site light travel time.  The latest event, GW170814,
was also detected at the Virgo site, in Italy.  The likelihood of
these being false-positive detections is reported as: less than
\mbox{$10^{-4}$~yr$^{-1}$} for GW150914, GW151226, GW170104 and
GW170814; less than \mbox{$3.3\times 10^{-4}$~yr$^{-1}$} for GW170608;
and 0.37~yr$^{-1}$ for LVT151012.  This paper examines the earliest
four events, GW150914, LVT151012, GW151226, and GW170104, using data
made available at the LIGO Open Science Center
(LOSC)~\cite{LOSC_2014}.

GW150914 had a sufficiently strong signal to stand above the noise
after removing spectral peaks and band-pass
filtering~\cite{PhysRevD.93.122004}.  For the other events, primary
signal detection involved the use of matched filters, in which the
measured strain records are cross-correlated with template waveforms
derived using a combination of effective-one-body, post-Newtonian and
numerical general relativity techniques~\cite{PhysRevLett.116.221101}.
Matched filter signal detection treats each template as a candidate
representation of the true gravitational wave signal and measures how
strongly the template's correlation with the measured signal exceeds
its expected correlation with detector noise.  Strong correlation
indicates a good match, but the true signal may not exactly match any
of the templates.  The dependence of matched filter outputs on
template parameters informs estimates of the physical parameters of
the source event~\cite{PhysRevD.78.124020,PhysRevLett.116.241102}.

Template-independent methods, using wavelets, were used to reconstruct
the waveform for GW150914 from the data~\cite{0264-9381-32-13-135012,
  PhysRevD.93.122004, PhysRevD.95.062002,PhysRevLett.116.241102},
achieving 94\% agreement with the binary black hole model.  A
template-independent search for generic gravitational wave bursts also
detected GW170104, but with lower significance than the matched filter
detection.  For GW170104 a morphology-independent signal model based
on Morlet-Gabor wavelets was used, following detection, to construct a
de-noised representation of the binary black hole inspiral waveform
from the recorded strain data~\cite{PhysRevLett.118.221101}.  This was
found to have an 87\% overlap with the maximum-likelihood template
waveform of the binary black hole model, which is statistically
consistent with the uncertainty of the template.

In~\cite{2016arXiv160206833T}, a Rudin-Osher-Fatemi total variation
method was used to de-noise the signal for GW150914, yielding a
waveform comparable to that obtained with a Bayesian approach
in~\cite{PhysRevLett.116.061102}.  The same authors also applied
dictionary learning algorithms to de-noise the Hanford signal for
GW150914~\cite{PhysRevD.94.124040}.

Cresswell et al~\cite{2017JCAP...08..013C} have independently analyzed
the LIGO data for GW150914, GW151226 and GW170104.  They report
correlations in the residual noise at the two sites\,---\,after
subtracting model templates obtained from the LOSC\,---\,and suggest
that a clear distinction between signal and noise remains to be
established.  Recognizing that a family of template waveforms may
``fit the data sufficiently well'', they claim that ``the residual
noise is significantly greater than the uncertainly introduced by the
family of templates''.  The claim of correlations in the residuals is
contrary to analysis in~\cite{PhysRevLett.116.221101}.  We discuss
below why we believe~\cite{2017JCAP...08..013C} is erroneous.

The present work introduces a new method for extracting signal
waveforms from the noisy strain records of black hole binary
coalescence events provided by the LOSC.  Noise that is inconsistent
with prior knowledge regarding timing and a reasonable-fit template
for an event is selectively filtered out to better reveal the
gravitational wave signal.  Our method relies on knowledge obtained
using the matched-filter techniques, discussed above, for signal
detection and identification of a reasonable-fit template.  We rely on
the (approximate) signal event times given by the LOSC and the broad,
time-dependent spectral features of the black hole coalescence
templates.\footnote{It is noted at https://losc.ligo.org/ that the
  provided numerical relativity template waveforms are consistent with
  the parameter ranges inferred for the observed events but were not
  tuned to precisely match the signals. ``The results of a full
  LIGO-Virgo analysis of this BBH event include a set of parameters
  that are consistent with a range of parameterized waveform
  templates.''  While a full LIGO-Virgo analysis may combine analyses
  of many templates, each consistent with the data, the variation of
  the time-dependent spectral features of the reasonable-fit templates
  will be inconsequential for the present work.}  In the case of
GW150914, we have also done signal extraction without using the
template, but assuming the event has the smoothly varying frequency
content typical of black hole inspiral, merger and ringdown.  Our
waveform extraction method does not use the templates' phase or
detailed amplitude information\,---\,instead such information is
obtained directly from the recorded data.  The extracted waveforms are
compared with similarly filtered templates to determine best-fit
amplitude and phase parameters and associated uncertainties.

Our limited objective in this work was to independently analyse data
provided by the LOSC in the hope of obtaining clean representations of
the black hole coalescence strain signals that could be compared with
the provided templates and published results.  Our analysis method is
\emph{not} designed to \emph{detect} gravitational wave events, and
should not be confused with the matched-filter techniques used for
such detection.  Application of our method to estimation of physical
parameters or to events other than black hole coalescence is beyond
the scope of the present work.

In Section \ref{S:Noise} we describe the characterization of detector
noise and identification and removal of spectral peaks due to AC line
power (60 Hz and harmonics), calibration signals and other
non-astrophysical causes.  Band-pass filters are used to remove the
high amplitude noise below about 30 Hz and at frequencies higher than
expected in the gravity wave events.  Statistical analysis gives no
indication that the filtered signals differ significantly from
band-limited Gaussian noise, with the exception of a few obvious
glitches.  Also, no significant correlation is found between
detectors.  Section \ref{S:Extraction} describes the use of
time-frequency bands to further reduce the influence of noise that
masks the astrophysical strain signals.  This allows determination of
the event time, phase and amplitude differences between the two
detectors, and construction of a coherent signal once the Hanford
signal time, phase and amplitude are adjusted to match Livingston
(which is taken as reference).  The clean signals are compared with
the reasonable-fit templates provided by the LOSC.  Analysis of
simulated events, with known signals injected into real noise,
demonstrates the reliability and uncertainties associated with our
signal extraction method.  In Section \ref{S:Discussion} we discuss
the relationship of our work to matched-filter signal detection,
provide some remarks on signal-to-noise ratios, and comment on
correlations of noise and residuals between detectors.  The final
section provides brief conclusions.

\section{Signal cleaning and noise characterization}\label{S:Noise}

The LOSC has made available time series records of the measured strain
data for each of the reported events.  The events are roughly central
within short (32~s) and long (4096~s) records, each available at 4096
samples per second (sps) and 16384 sps.  The 4096 sps
records\,---\,used in the present work\,---\,were constructed from the
16384 sps records by decimation.

\begin{figure*}[t!]
  \center\includegraphics[height=2.25in]{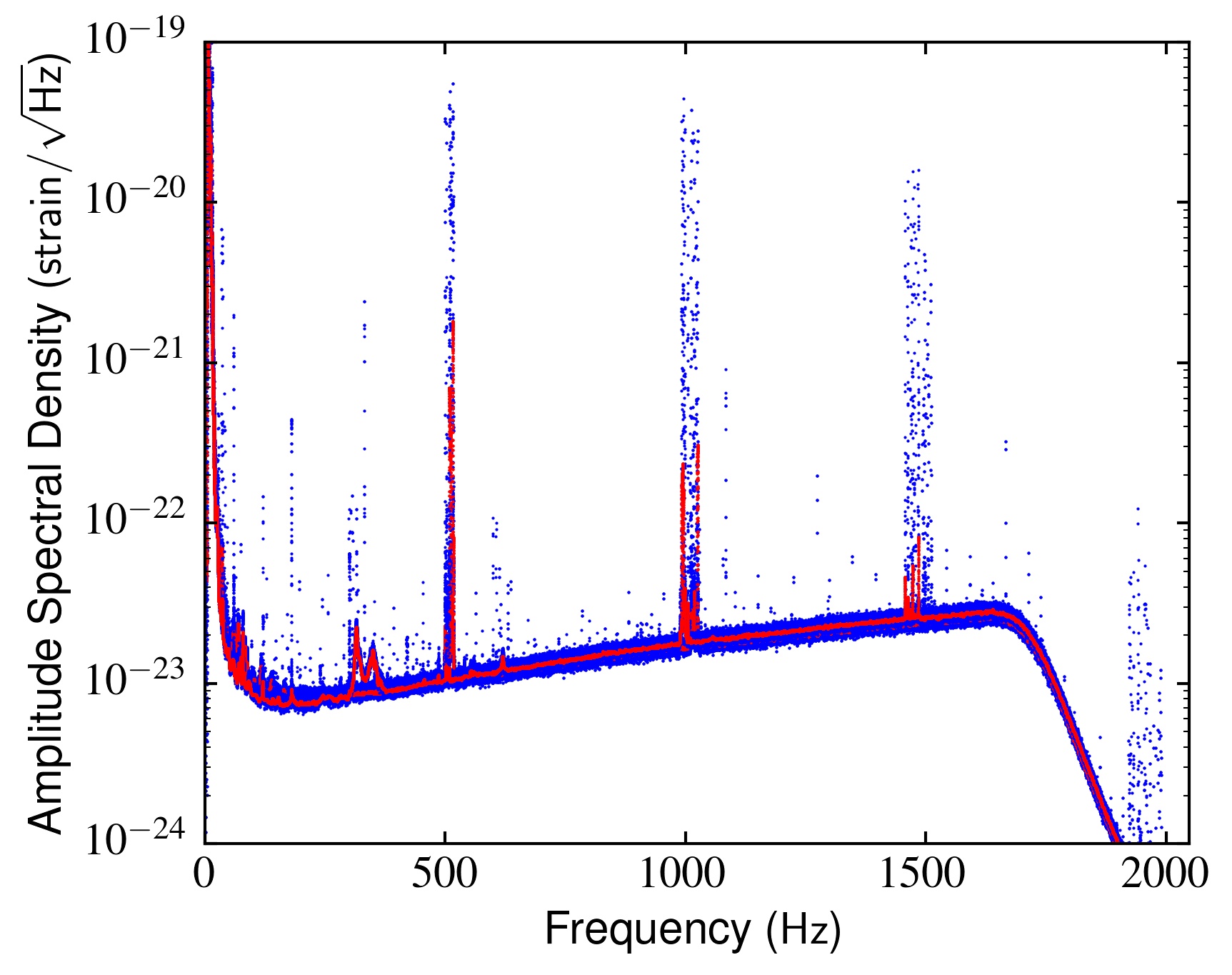}\hspace*{1.5cm}\includegraphics[height=2.25in]{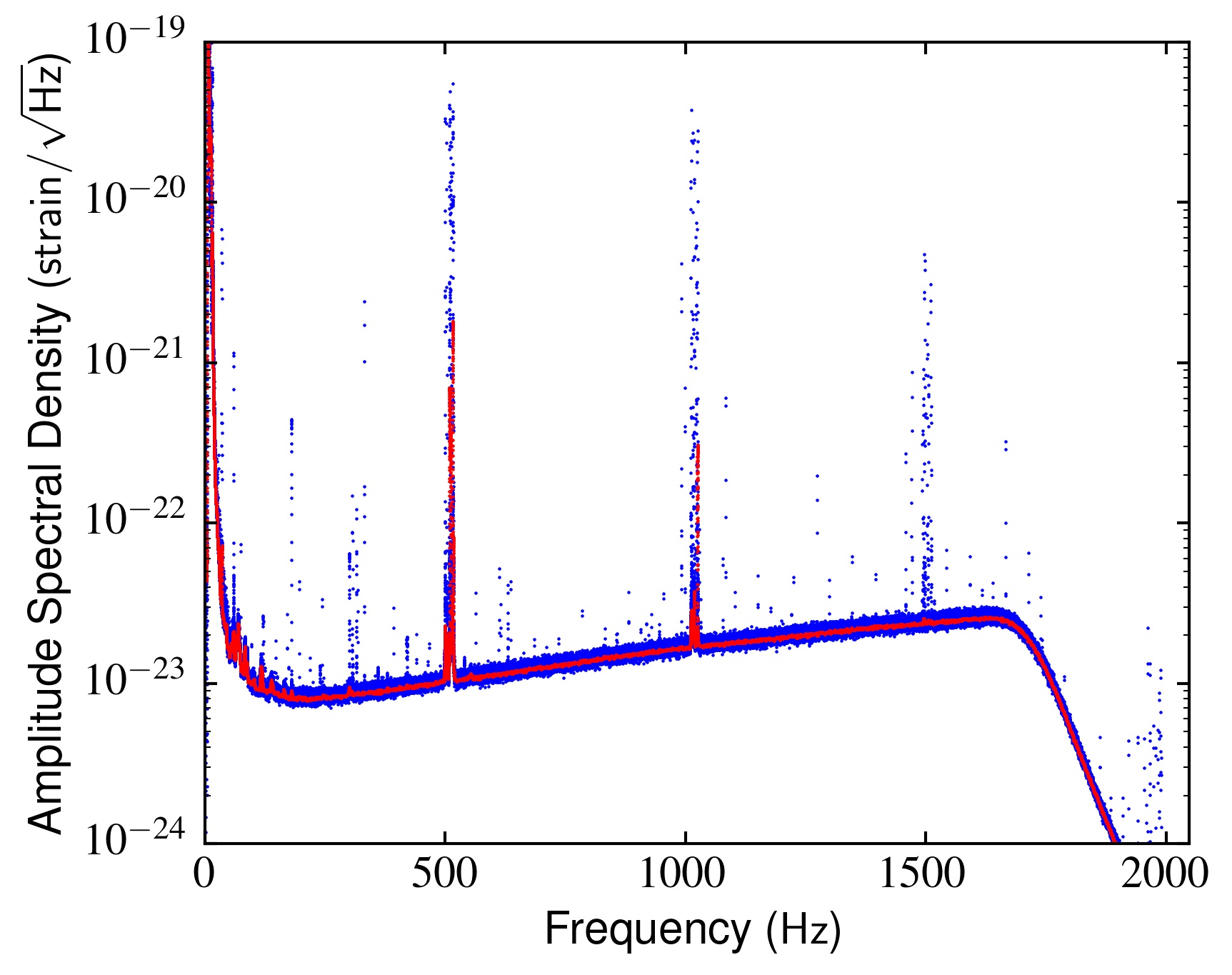}
  \caption{Amplitude spectral density
    ($\mathrm{ASD}\equiv\sqrt{S(f)}$) of detector noise (blue) and the
    smoothed baseline $\sqrt{S_b(f)}$ (red) from 4096 s strain records
    at Hanford (left panel) and Livingston (right panel) at the time
    of GW150914.}
  \label{F:ASD}
\end{figure*}

From each long strain record, $s_l(t)$, the power spectral density
(PSD), $S(f)$, was constructed using Welch's average periodogram
method with overlapping 64 s segments and Planck
windows~\cite{0264-9381-27-8-084020} with 50\% tapers.  This enabled
identification of the positions and widths of numerous sharp spectral
peaks corresponding to AC powerline harmonics, detector calibration
signals, and other deterministic sources.  A smoothed PSD baseline,
$S_b(f)$ was constructed, corresponding to the PSD with the narrow
peaks removed.  This was used as the PSD for subsequent analysis of
the given event and detector, and for whitening of signals. The square
roots of the PSDs and baselines (i.e., the amplitude spectral
densities (ASDs)) found for GW150914 are shown in Figure \ref{F:ASD}.

\begin{figure*}[t!]
  \center\includegraphics[height=2.25in]{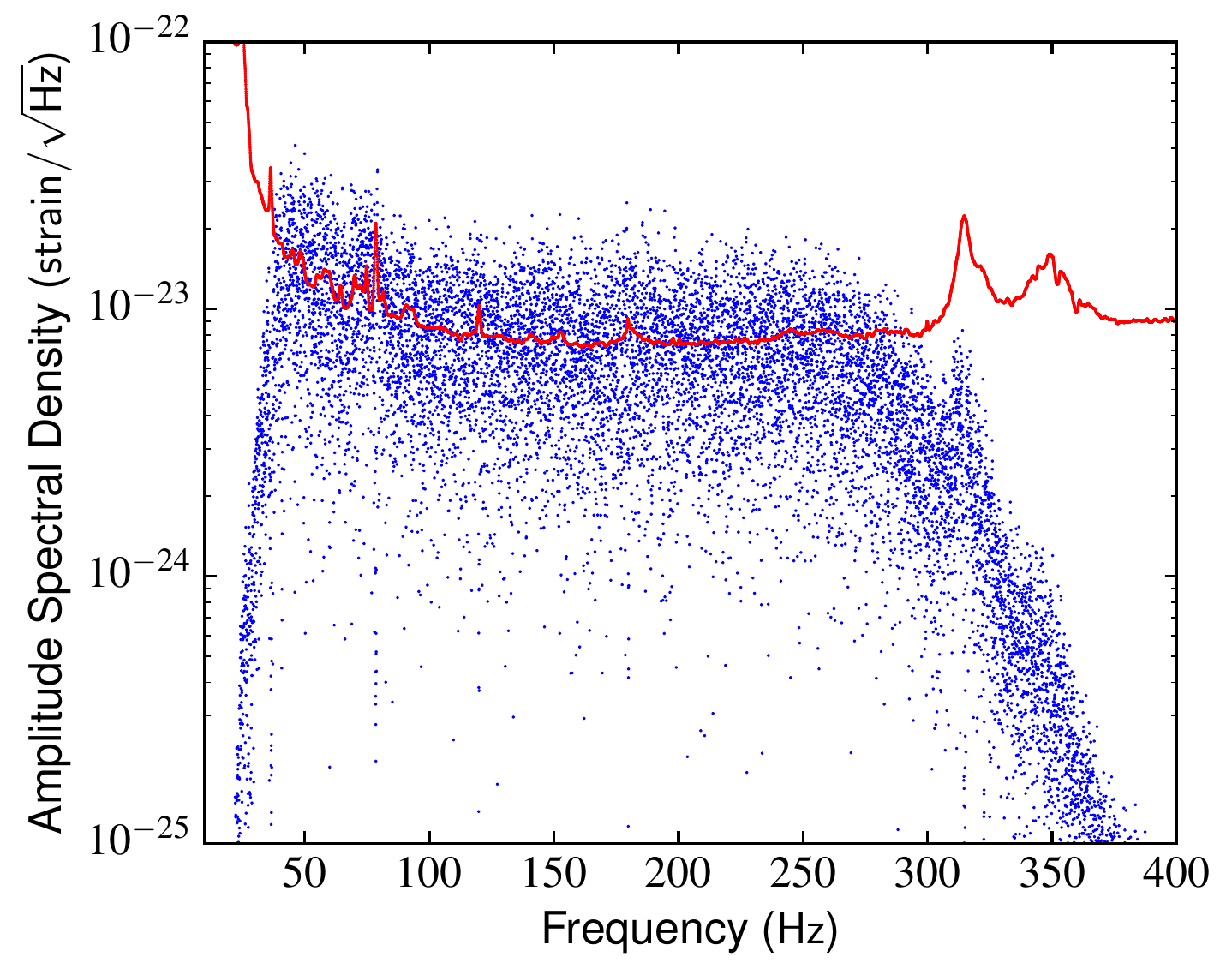}\hspace*{1.5cm}\includegraphics[height=2.25in]{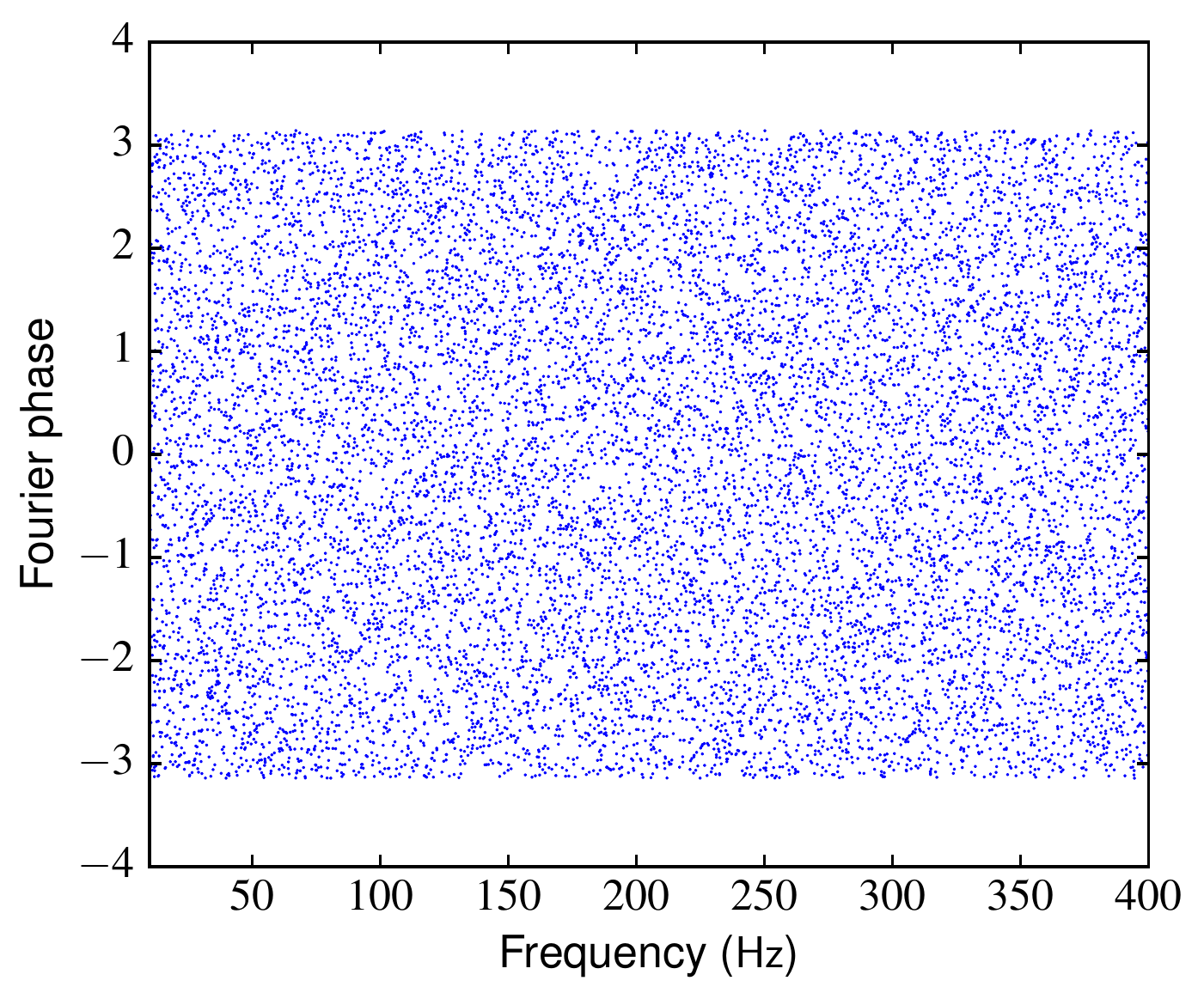}
  \caption{Left panel: normalized magnitude of $\tilde{s}_{cbp}$
    (blue) and ASD baseline $\sqrt{S_b(f)}$ (red) for Hanford at the
    time of GW150914.  Right panel: phase of $\tilde{s}_{cbp}$.  (A
    plot of phase differences of adjacent frequencies looks equally
    random.)}
  \label{F:FFT}
\end{figure*}

\begin{figure*}[t!]
  \center\includegraphics[height=2.25in]{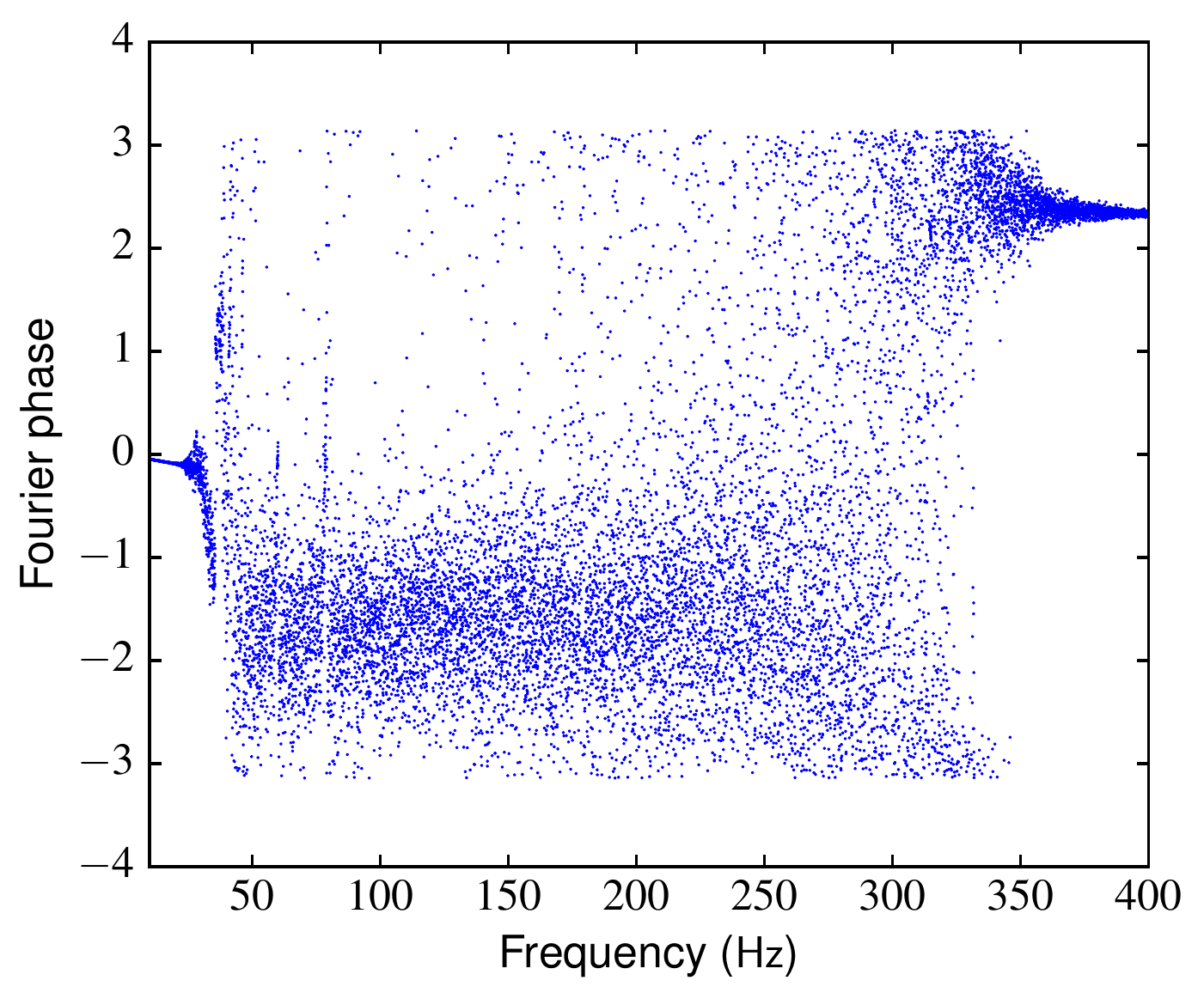}\hspace*{1.5cm}\includegraphics[height=2.25in]{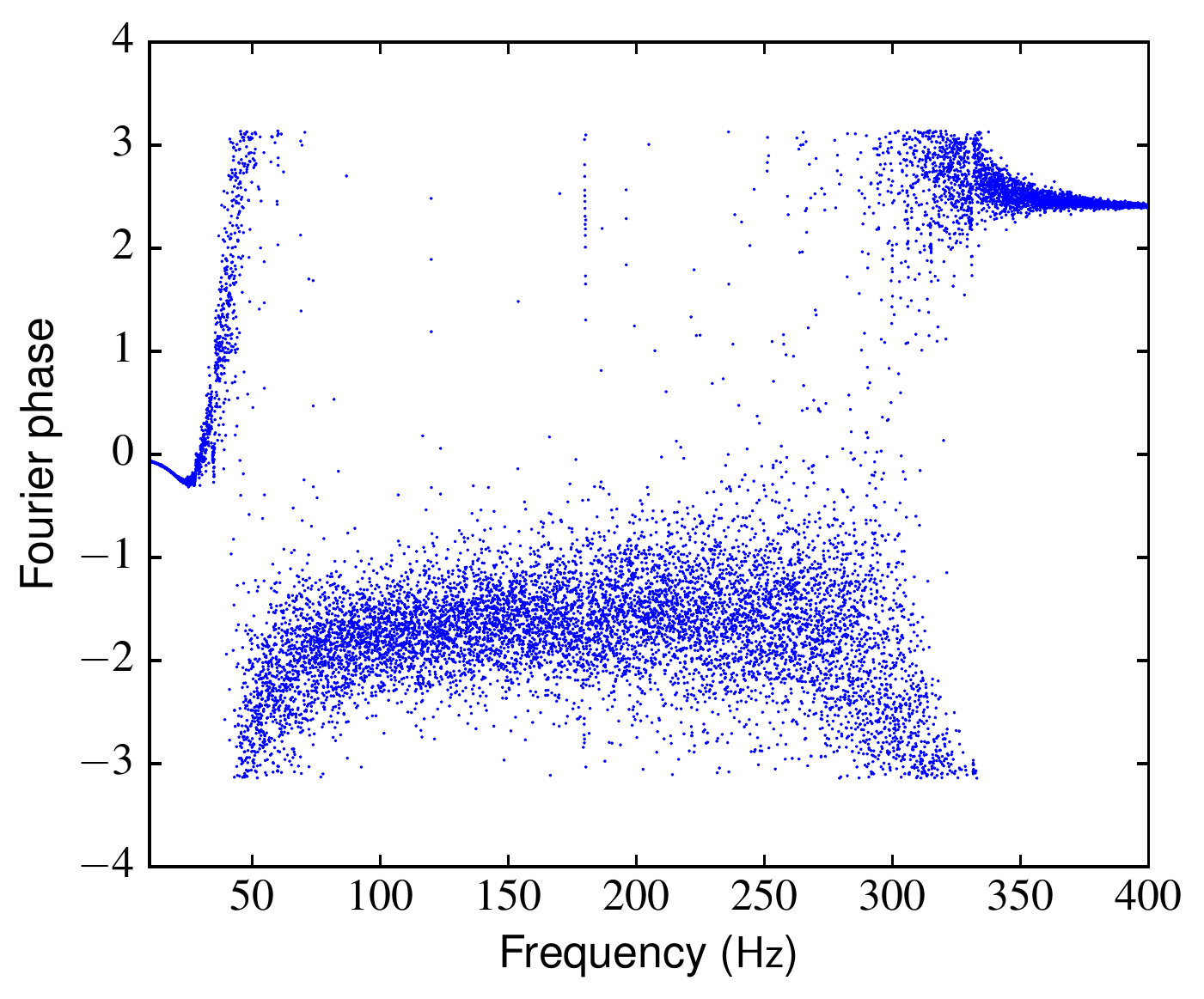}
  \caption{Phase of $\tilde{s}_{cbp}$ for Hanford (left) and
    Livingston (right) when filtering was performed without using
    window functions to prevent border distortion.}
  \label{F:badPhase}
\end{figure*}

Order 2 Butterworth notch filters were used to remove the
deterministic signals from the 32 s records, $s(t)$, leaving a clean
signal, $s_{clean}(t)$, in the pass bands of interest.  The filter
frequencies and widths were manually adjusted to reduce obvious
spectral peaks to the level of the broadband noise.  A window with
0.5~s Hann end tapers was used to avoid border distortion.  Throughout
this work, time-domain filters were applied forward and backward to
nullify any phase change.  Finally, noise outside the band of interest
was suppressed by a band pass filter, yielding $s_{cbp}(t)$.  The pass
band was adjusted for each event to optimize the signal extraction,
with extreme bounds of 35~Hz and 315~Hz for the studied events.
Figure \ref{F:FFT} shows the magnitude and phase of $\tilde{s}_{cbp}$
for GW150914 at Hanford, where $\tilde{s}$ is the discrete Fourier
transform of $s$ and the amplitude is normalized to match the total
power of $s_{cbp}$.  Equivalent plots for Livingston and for the other
events are very similar.

\begin{figure*}[t]
  \center\includegraphics[height=2.25in]{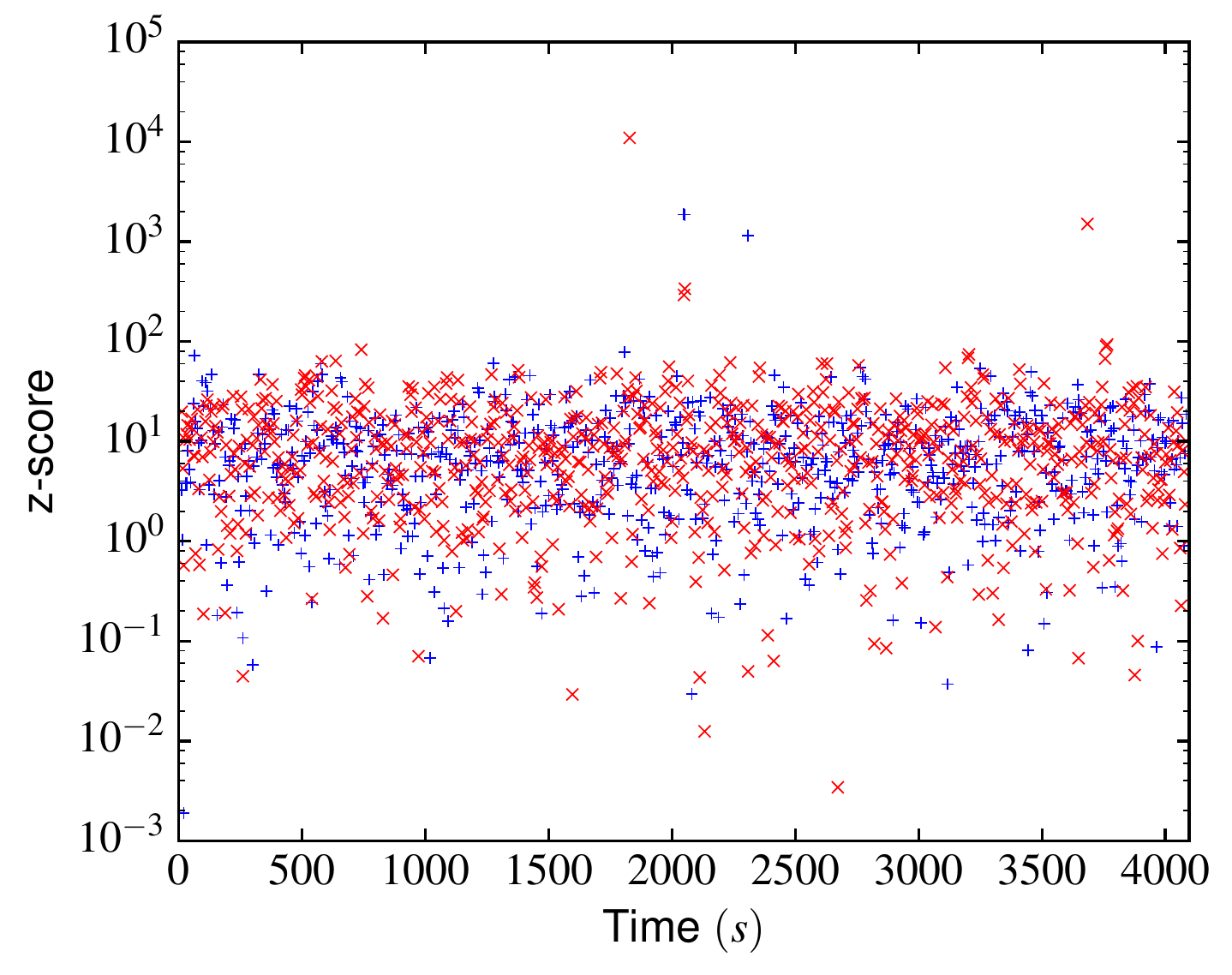}\hspace*{1.5cm}\includegraphics[height=2.25in]{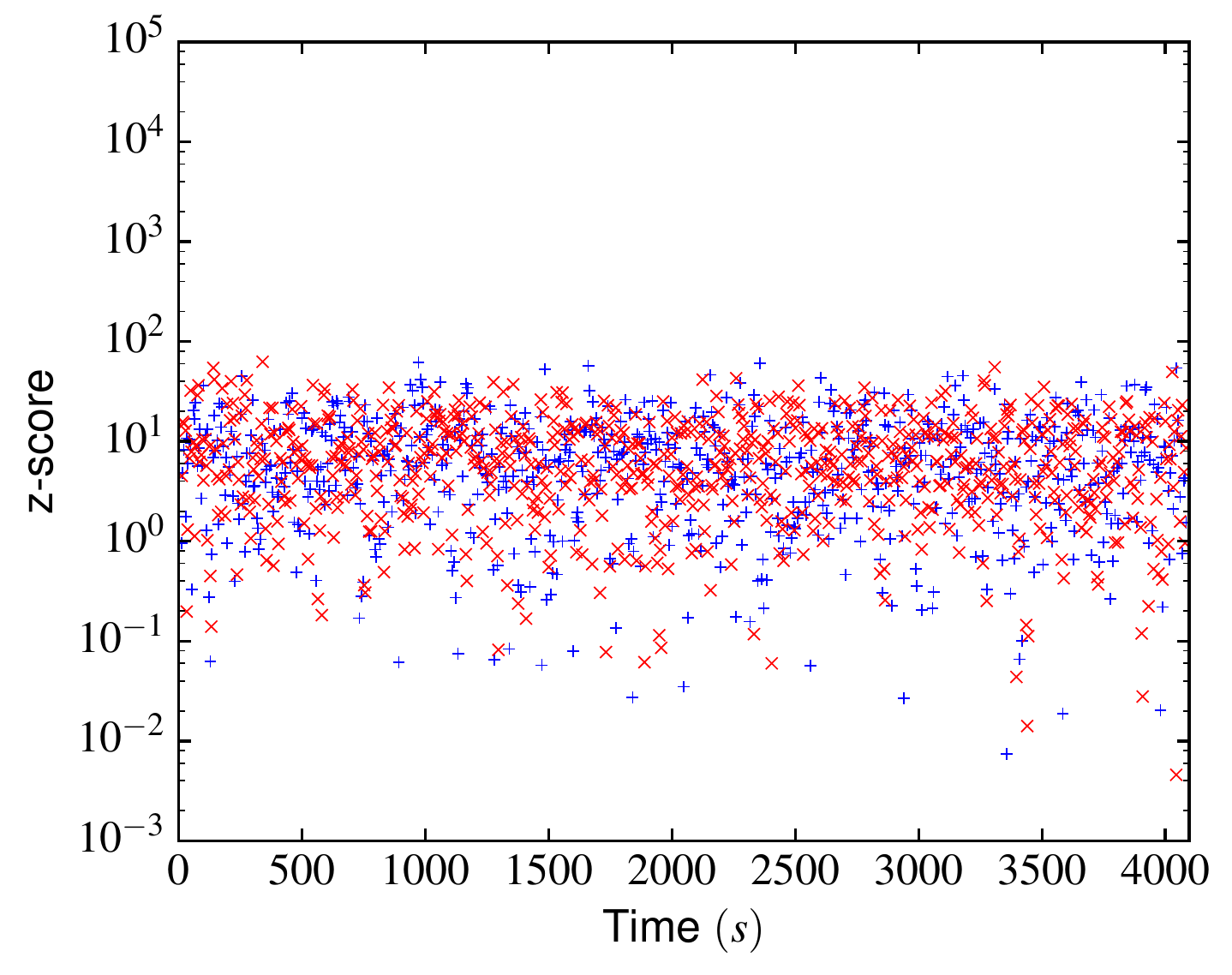}
  \caption{Left panel: GW150914 - z-scores from normaltest of measured
    noise for Hanford (blue $+$) and Livingston (red $\times$).  Right
    panel: equivalent z-scores for band-passed Gaussian noise.}
  \label{F:z-scores}
\end{figure*}

Doing the above signal cleaning without the use of smoothly tapered
window functions to avoid border distortion gives similar spectral
amplitudes, but the phase plots, shown in Figure \ref{F:badPhase}, are
no longer random.  Comparison of these latter plots with similar plots
in ref.~\cite{2017JCAP...08..013C} reveals great similarity.  It thus
appears that the authors of~\cite{2017JCAP...08..013C} may not have
used suitable window functions.  The effects of border distortion are
also apparent elsewhere in~\cite{2017JCAP...08..013C}, to the extent
that the reliability of the findings and conclusions of that work must
be questioned.

Although we do not give the specifics in every case, window functions
have been applied wherever warranted in the work reported here.  For
each situation, window parameters have been chosen to avoid border
distortion effects while minimizing bias of useful information.

Statistical tests were performed on the long records for each of the
four events.  First, 255 overlapping 32~s records were drawn from each
of the long records.  These short records were cleaned, band-pass
filtered, and whitened by scaling in the frequency domain by
$\sqrt{N/S_b(f)}$ (where $N$ is arbitrarily chosen as the mean of
$S_b(f)$ in the 3rd quartile of the pass band).  Then three
overlapping 8~s records were drawn from the central 16~s of the
processed 32~s records, avoiding window effects on the ends of the
32~s records.  Each 8~s record was tested for normality using the
z-score of D'Agostino and Pearson, which gives a combined measure of
skew and kurtosis~\cite{10.2307/2335012,
  doi:10.1080/00031305.1990.10475751, 10.2307/2335960}.  Figure
\ref{F:z-scores} shows the z-scores for all the 8~s records drawn from
the GW150914 data.  Also shown is a plot where the data was randomly
generated with a normal distribution, band-pass filtered with the same
pass band as the real data (and using the same window function), and
then subjected to the same `normaltest'.  High z-score values from the
measured data correspond to obvious glitches in one detector or the
other, and to the detected event at both detectors.  Otherwise, the
measured and randomly generated data have similar z-scores.

\begin{figure}[h]
  \center\includegraphics[height=2.25in]{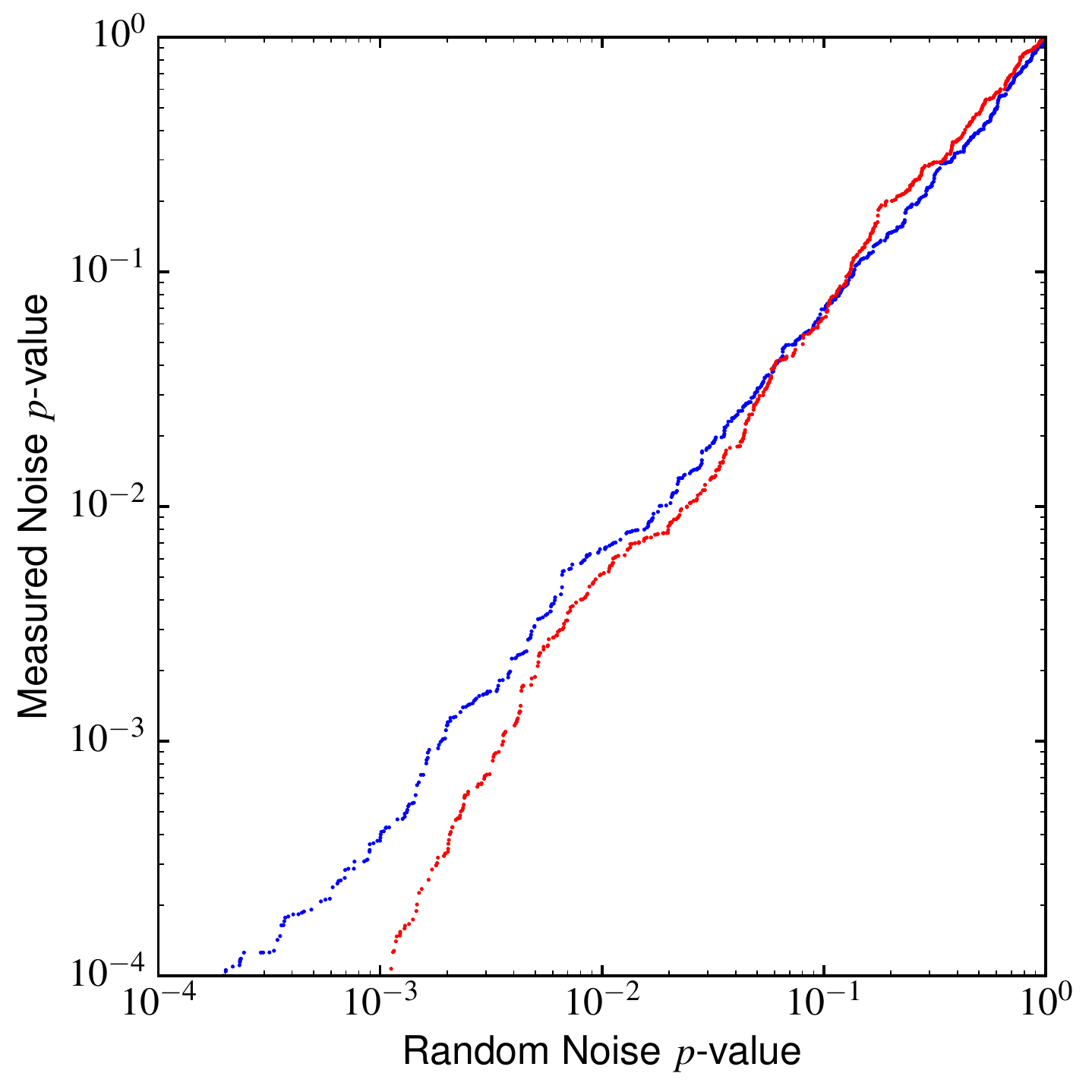}\\
\includegraphics[height=2.25in]{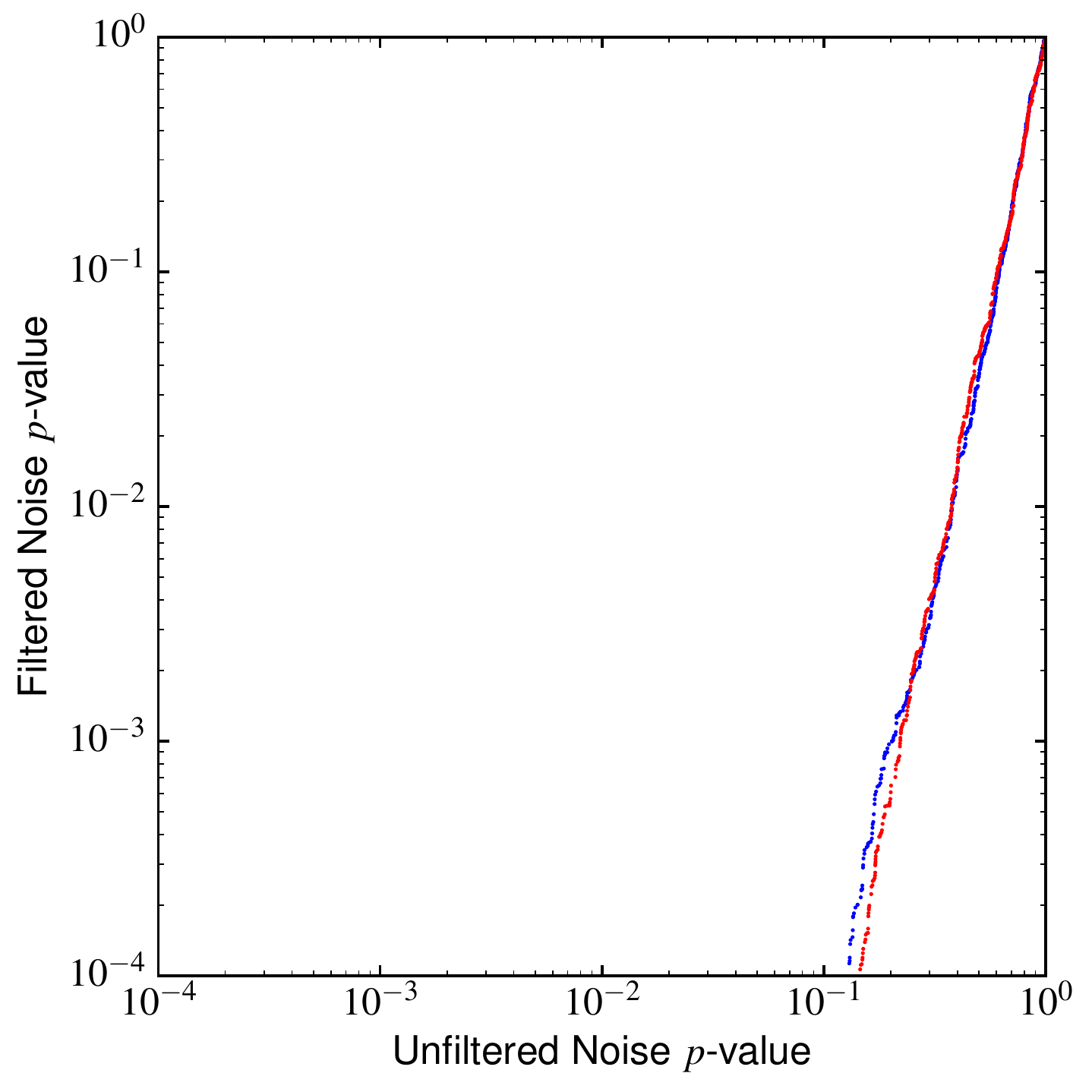}
\caption{Top panel: sorted $p$-values of cleaned, band-passed and
  whitened noise records from Hanford (blue) and Livingston (red),
  plotted against sorted $p$-values of similarly band-passed Gaussian
  noise records.  Each $p$-value estimates the likelihood that the
  32768 samples in a record have been generated by a Gaussian random
  process.  Bottom panel: a similar plot showing $p$-values for
  band-passed versus unfiltered Gaussian noise records.}
  \label{F:p-values}
\end{figure}

The normaltest also generated $p$-values for the null hypothesis that
the data comes from a normal distribution.  As the number of normally
distributed samples in a record becomes infinite, the $p$-value should
go to 1.  But a set of finite length records, even if all samples are
generated by a Gaussian (pseudo-)random process, will have a
distribution of $p$-values less than 1.  Figure \ref{F:p-values}
compares the $p$-value distributions for the filtered and whitened
measured strain records and the filtered, randomly generated records.
The lower panel demonstrates that unfiltered records of Gaussian
random noise tend to have significantly higher $p$-values than similar
records that have been band-pass filtered in the same way as the
GW150914 data.  The (filtered) measured data is just slightly less
likely to be identified as Gaussian than is the band-passed Gaussian
noise, with Livingston having somewhat larger reductions than Hanford.

\begin{figure}[h]
  \center\includegraphics[height=2.25in]{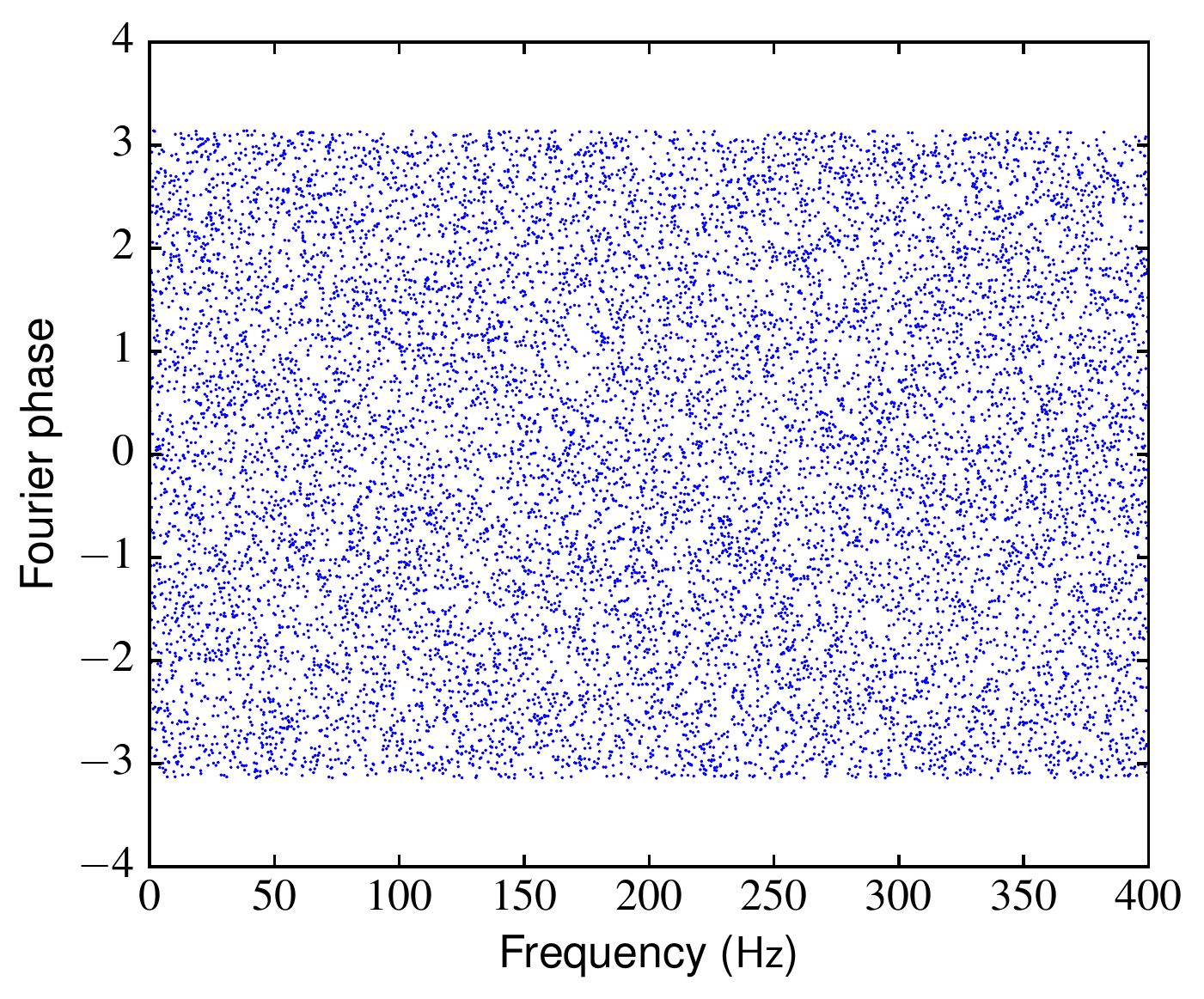}
  \caption{Phase from Fourier transform of the sum of
    cross-correlations between Hanford and Livingston for 255
    overlapping, cleaned and filtered 32~s records drawn from each of
    the 4096~s records for GW150914.}
  \label{F:CCphase}
\end{figure}

A final test looked for correlations between the two detectors.  The
cross-correlations between detectors for all 255 overlapping, cleaned and
filtered 32~s records drawn from the 4096~s record for GW150914 were
summed.  Fig. \ref{F:CCphase} shows phases from the Fourier transform
of this composite cross-correlation; there are no significant features
in the magnitude spectrum.  The apparent randomness of the phases is
consistent with the absence of correlations between the broadband
noise at the two detectors.

\section{Signal extraction}\label{S:Extraction}

The techniques used by LIGO to detect and characterize black hole
coalescence signals tell us at what time the events appear in the data
records $s(t)$ and yield template waveforms
$H(t)=\mathfrak{h}_p(t)+i \mathfrak{h}_c(t)$ that fit the observations
reasonably well.\footnote{Templates were obtained from the file
  LOSC\_Event\_tutorial.zip at https://losc.ligo.org/, on 2018/07/18.
  The complex template $H$ is used for convenience.  A real template
  is obtained by combining $\mathfrak{h}_p$ and $\mathfrak{h}_c$ in
  quadrature, using amplitudes that best fit observations.}  In the
following sub-sections we describe how, using the approximate event
time and approximate time-dependent frequency content of the template,
we extract from the records $s(t)$ much cleaner representations
$s_f(t)$ of the gravitational strain waveforms and the associated
uncertainties.  (The full template waveform is used at the outset to
determine the approximate signal time offset between sites, but
subsequent iterative analysis obviates influence of the template's
detailed phase and amplitude information.)  We also describe how our
method is adapted, in the case of a strong signal like GW150914, to
obtain a clean strain waveform without using a template, but simply
assuming the signal has the smoothly varying frequency content
characteristic of a black hole inspiral, merger and ringdown.

The relation of the cleaned strain signals at the two sites to each
other and to the template is characterized as follows.  Best-fit
values are obtained for the difference of signal arrival times:%
\beq%
\Delta t_{event}=t^H_{event}-t^L_{event}\,,%
\eeq%
where the superscripts $H$, $L$ refer to the two detectors.  The
complex templates are fitted to the extracted waveforms to determine,
for each detector $D$, the best-fit relative amplitude $A^D$ and phase
$\phi^D$ for the real-valued template:%
\beq%
h^D=A^D\,(\mathfrak{h}_p\,\cos{\phi^D}-\mathfrak{h}_c\,\sin{\phi^D})\,.\label{Eq:template}%
\eeq%
Unless needed for clarity, the superscripts $D$, $H$, $L$ will often
be omitted below.  The phase difference $\Delta\phi=\phi^H-\phi^L$ is
also independently determined, without using knowledge of $\phi^H$ or
$\phi^L$.\footnote{The LIGO-provided $(\mathfrak{h}_p,\mathfrak{h}_c)$
  are given with amplitudes expected for a coalescing source system
  with an effective distance of 1~Mpc.  An observed source with
  relative amplitude $A$ then has the approximate effective distance
  $d_{eff}\approx 1/A$~Mpc.  A more accurate distance calculation
  depends on $A^H$, $A^L$, $\Delta\phi$, and other factors.}

The extracted waveforms may, and indeed do, differ from the templates.
Differences between the signals extracted from the two detectors are
used to estimate the residual noise.  Statistical analysis supports
determination of uncertainties in time offsets, template amplitudes
and phases, and residual differences between the extracted signals and
the templates.

\subsection{Extraction process}

\begin{figure*}[t]
  \center\includegraphics[height=3.25in]{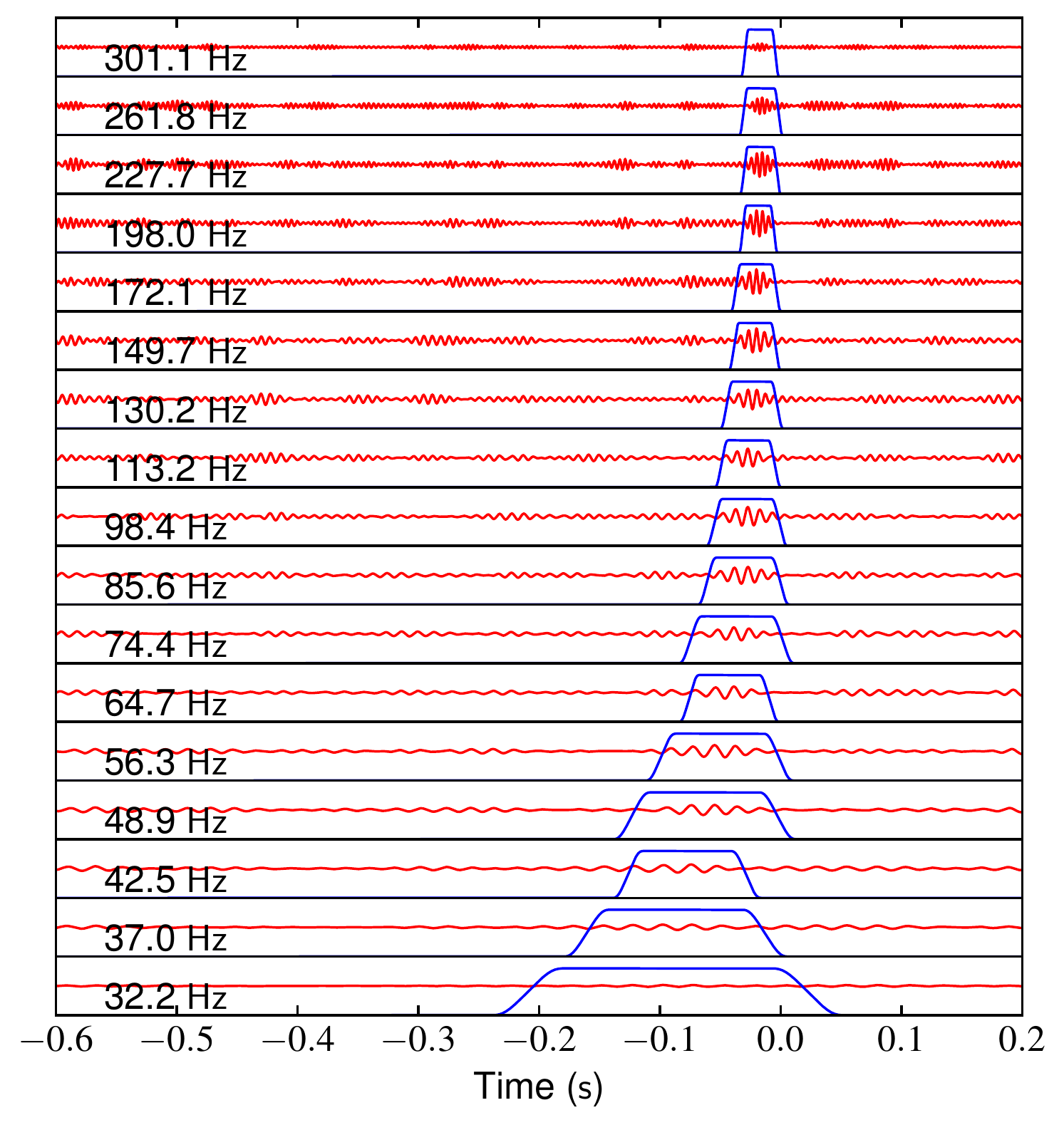}\hspace*{1.5cm}\includegraphics[height=3.25in]{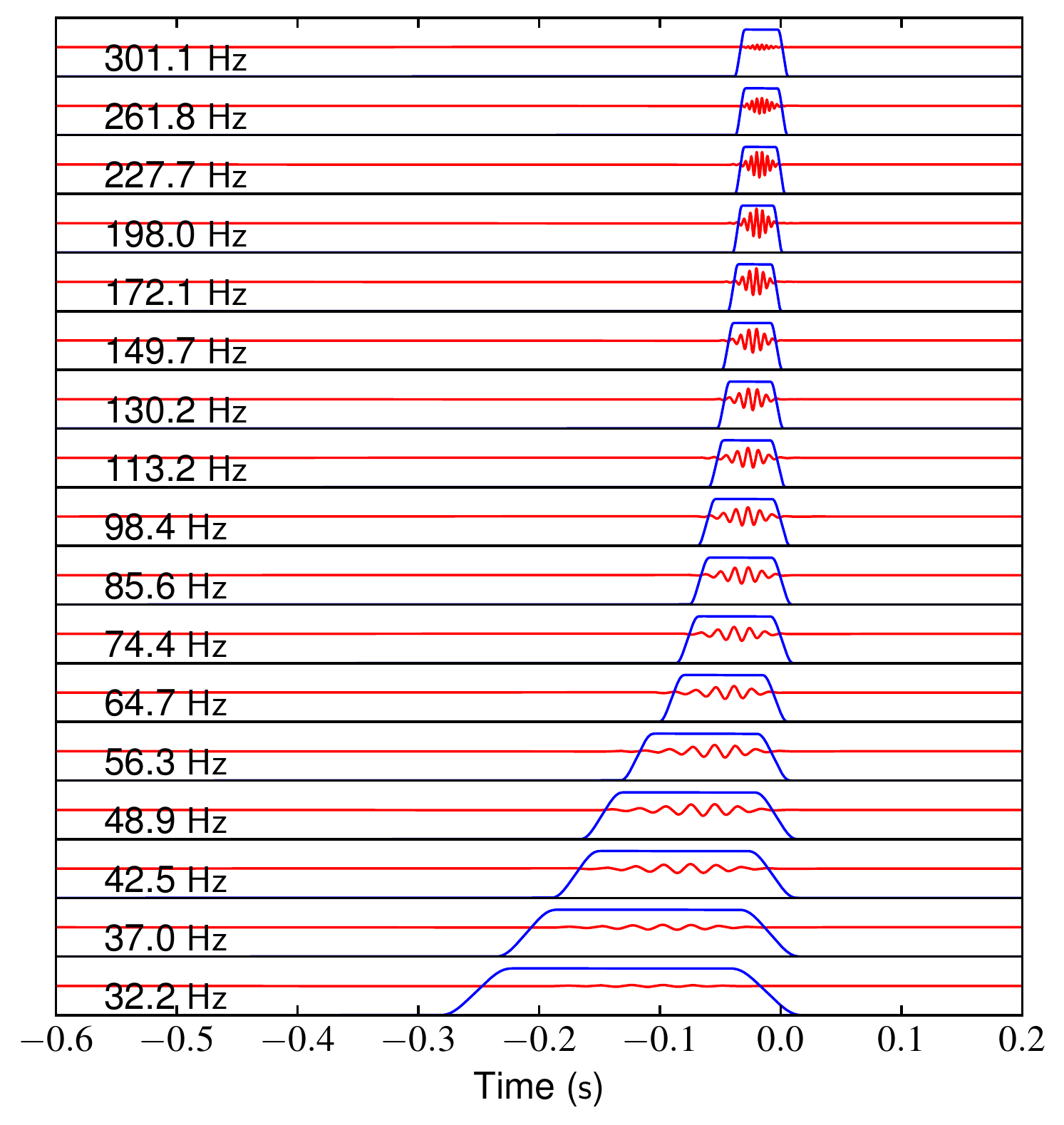}
  \caption{Time-frequency bands for GW150914 derived without use of
    the template (left) and using the template (right).  The red
    waveforms are the filtered signal $\mathcal{F}_{b_i}(s^L_{cbpw})$
    and template $\mathcal{F}_{b_i}(\mathrm{Re}(H_{cbpw}))$,
    respectively, at the indicated frequencies, $f_i$.  The
    corresponding windows $W_i$, with Planck tapers, are in blue.}
  \label{F:timeFreq}
\end{figure*}
\begin{figure}[h!]
  \center\includegraphics[height=3.25in]{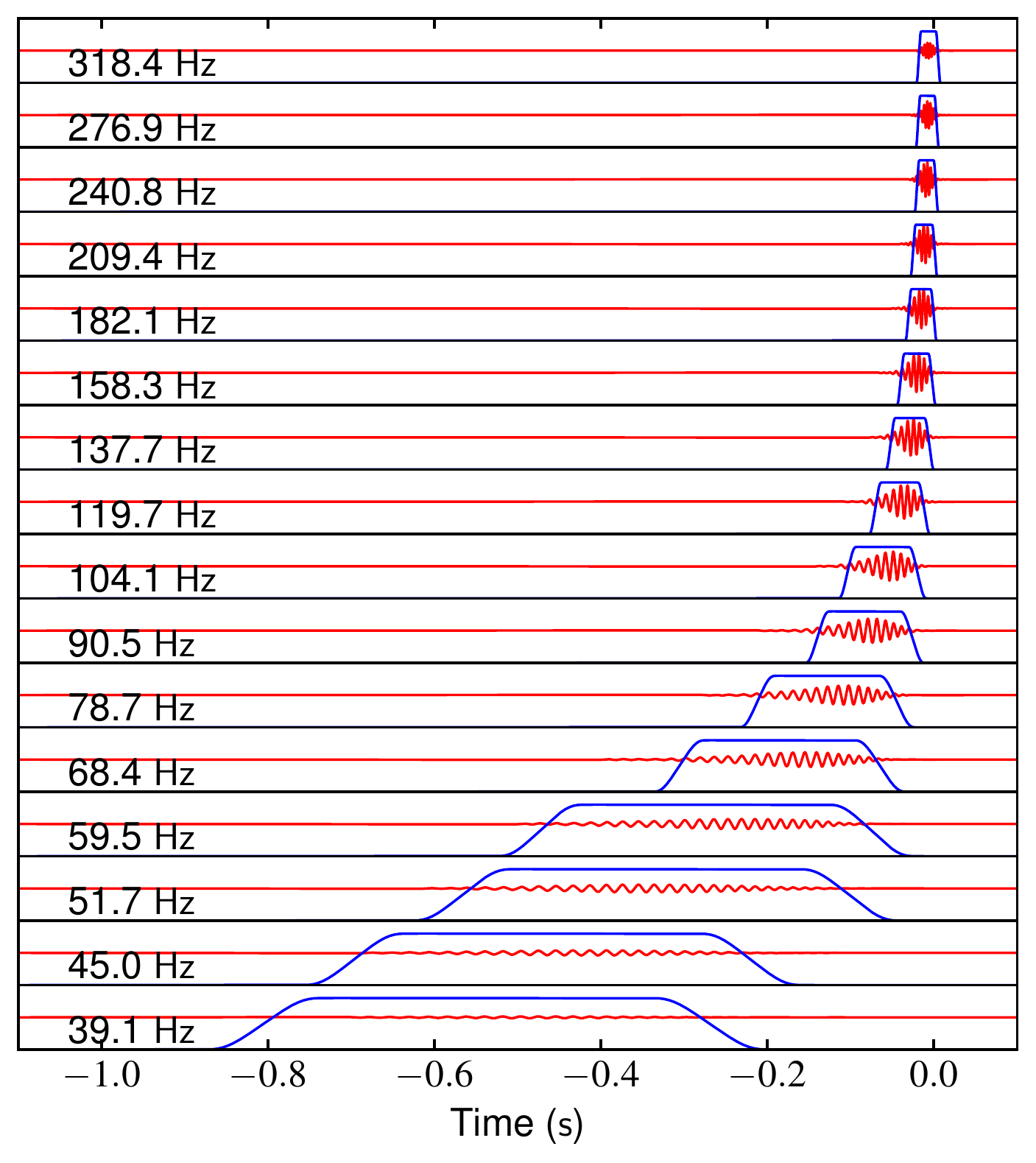}
  \center\includegraphics[width=3.15in]{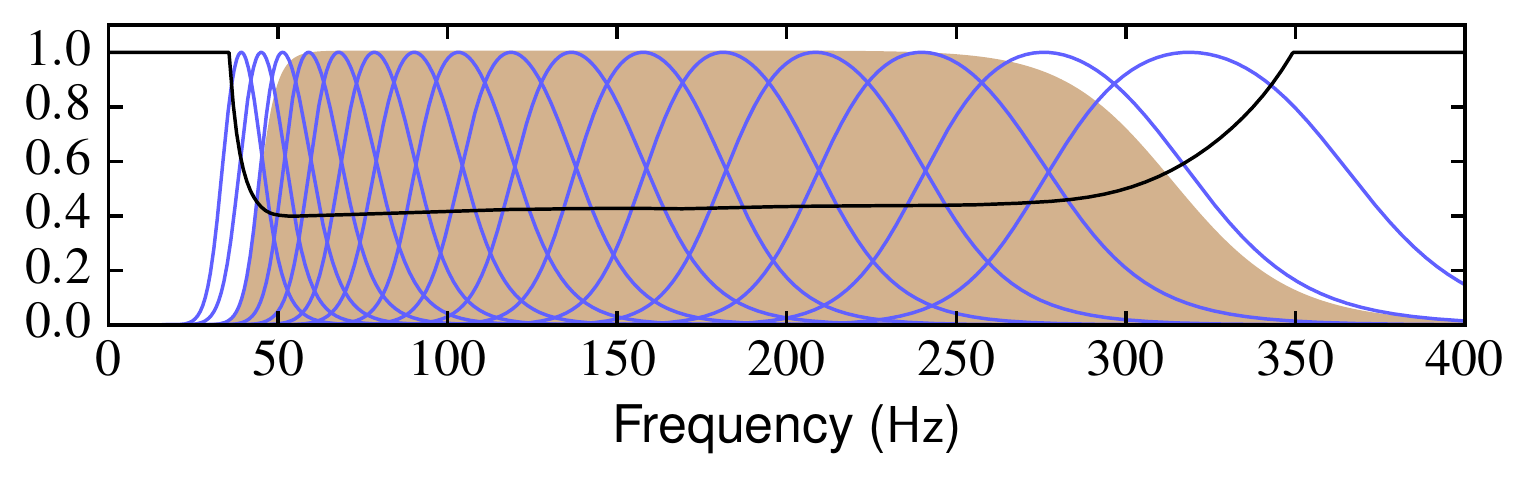}
  \caption{Top panel: Similar to the right panel of
    Fig.~\ref{F:timeFreq}, for GW151226.  Bottom panel: Filter
    response amplitudes for GW151226.  The light brown background shows the
    \mbox{45 -- 315 Hz} pass band; the blue lines show the narrow
    bands of the top panel; and the black line the normalization
    $1/N(f)$ that compensates for band overlaps.}
  \label{F:timeFreq2}
\end{figure}

The signal filtering described in Section \ref{S:Noise} eliminates
deterministic noise (i.e., spectral peaks) and noise outside the
frequency pass band ultimately chosen to best reveal the signal.  For
the resulting filtered 32~s records, we write%
\beq%
s_{cbp}(t)=n(t)+g(t)\,,%
\eeq%
where $n(t)$ is band-limited, nearly Gaussian noise and $g(t)$ is a
filtered representation of the true gravitational wave at the
detector.  We expect $g$ to have time-dependent frequency content
similar to $H_{cbp}$ (which is the template $H$ filtered the same as
$s$ was to obtain $s_{cbp}$).  Exploiting this similarity, our
strategy is to implement additional filters that selectively reduce
all portions of $s_{cbp}$ that are incompatible with the
time-dependent frequency content of $H_{cbp}$.  This minimizes $n$ to
better reveal $g$.  In the following, we have arbitrarily chosen
Livingston as the reference site.  Choosing Hanford, instead, would
not have materially changed the results.

Iterative calculations, with too many details to describe here, were
necessary to converge on best-fit signal time offsets, amplitudes and
phases and extract corresponding signals $s_f$.  Schematically, the
steps followed are:\vspace*{3pt}

\noi\textbf{(i)~\emph{Reduce time-span}:} Select, for use in all
subsequent steps, 4 seconds of cleaned, band-passed data starting
2.8~s before the nominal event time.  Do this for the template also.
All of the studied events have duration and usable bandwidth that
ensures observations outside this time window will contribute only to
$n(t)$.\vspace*{3pt}

\noi\textbf{(ii)~\emph{Synchronize signals}:} Cross-correlate the
filtered template $H^D_{cbp}$ and signal $s^D_{cbp}$, for each $D$, to
find the time of best match.  Interpolate, using a quadratic fit in
the neighborhood of the peak, to achieve time resolution finer than
the $1/4096$~s data time steps.  Using $s^L_{cbp}$ as reference,
time-shift $s^H_{cbp}$ and $H^D_{cbp}$ so all these signals become
coincident.  For GW150914, cross-correlation between $s^H_{cbp}$ and
$s^L_{cbp}$ was sufficient to determine their relative time offset.

(Synchronization was also done using whitened waveforms $s_{cbpw}$ and
$H_{cbpw}$ instead of $s_{cbp}$ and $H_{cbp}$, where the whitening was
performed in the frequency domain by multiplying by $\sqrt{N/S_b(f)}$
as described in Section \ref{S:Noise}.  Differences between the
results with and without whitening were not significant.)\vspace*{3pt}

\noi\textbf{(iii)~\emph{Choose time-frequency (t-f) bands}:} Define
overlapping, narrow frequency bands%
\beq%
b_i=(f_i/1.15,f_i\times 1.15),~~f_{i+1}=f_i\times 1.15%
\eeq%
that span the frequency pass band chosen for the event.  The ratio
1.15 was chosen to divide the signal into about 16 bands.  No attempt
was made to optimize this ratio\,---\,decreasing it would increase the
computational cost, while increasing it significantly would result in
less effective noise rejection.  Using each $b_i$, apply a Butterworth
band pass filter $\mathcal{F}_{b_i}$ to $H_{cbpw}$ (or, for
template-independent analysis, to $s_{cbpw}$).  In simplified form,
the algorithm for choosing the time windows is as follows.  For each
$i$, find the contiguous time window $W_i$ surrounding the maximum of
the envelope of $\mathcal{F}_{b_i}(H_{cbpw})$ (or of the
autocorrelation of $\mathcal{F}_{b_i}(s_{cbpw})$) and extending to
$1/2$ of the maximum (or an inflection point).  Symmetrically scale
the width of $W_i$ by the parameter $\alpha$, which is tuned for each
event.  (When using $s_{cbpw}$, time-shift $W_i$ so it encloses the
peak of the envelope of $\mathcal{F}_{b_i}(s_{cbpw})$.)  Finally, add
a Planck taper to each end of $W_i$, further increasing the total
non-zero window length by 50\%.

Choosing narrower time windows rejects more noise at the expense of
increased signal distortion\,---\,this proved beneficial for events
with weak signals.  The values chosen for $\alpha$ were $1.7$ for
GW150914 and GW170104, $1.3$ for LVT151012, and $1.215$ for GW151226.
We discuss $\alpha$ further in Section~\ref{S:Discussion}.

Figure \ref{F:timeFreq} shows the resulting windows $W_i$ for GW150914
computed without and with use of the template. The upper panel of
Figure \ref{F:timeFreq2} shows the $W_i$ for GW151226, computed using
the template.\vspace*{3pt}

\noi\textbf{(iv)~\emph{Apply t-f bands}:} For each $i$, and each
detector, construct the narrowly filtered waveform
$s_i=\mathcal{F}_{b_i}(s_{cbp}\times W_i)$, and its Fourier transform
$\tilde{s}_i$.  Sum the $\tilde{s}_i$ to give a composite
frequency-domain representation:%
\beq%
\tilde{s}_f=\frac{1}{N(f)} \sum\limits_i{\tilde{s}_i}\,,%
\eeq%
where $1/N(f)$ compensates for overlapping bands.  The lower panel of
Figure \ref{F:timeFreq2} shows the amplitude response functions of the
individual filters $\mathcal{F}_{b_i}$ for GW151226 and the
corresponding $1/N(f)$.  Transform back to the time domain to obtain
the filtered signals $s^H_f$ and $s^L_f$.  Apply the \emph{t-f} bands
to the template in the same way to obtain $H^H_f$ and $H^L_f$, where
the $H$-$L$ distinction arises due to the different spectral peaks at
the two detectors.\vspace*{3pt}

\noi\textbf{(v)~\emph{Match phases and amplitudes}:} Having already
made $s^H_f$, $H^H_f$ and $H^L_f$ synchronous with $s^L_f$, find, for
each detector, the phase $\phi^D$ and amplitude $A^D$ that yield from
$H^D_f$ a real template $h^D_c$ that best matches $s^D_f$.  (See
Eq.~(\ref{Eq:template}).)  The desired phase is the angle in the
complex plane at which the cross-correlation of $s^D_f$ and $H^D_f$
has maximum amplitude.  As done for precise time offsets, interpolate
to determine a precise phase.  The unscaled, phase-coherent real
template is then%
\beq%
h^D_f=Re\left(H^D_f \exp(i\phi^D)\right)\,.%
\eeq%
Finally, $h^D_c=A^D\,h^D_f$, where $A^D$ is given by%
\beq%
A^D=\frac{s^D_f\cdot h^D_f}{h^D_f\cdot h^D_f},%
\eeq%
with $a\cdot b=\sum{a_i b_i}$ the inner product of the two waveforms.

The template-dependent phase and amplitude relating $s^H_f$ to $s^L_f$
are given by $\Delta\phi=\phi^H-\phi^L$ and $A^{LH}=A^L/A^H$.  Use
these to obtain $s^H_c=A^{LH}R(s^H_f,\Delta\phi)$, where $R(s,\phi)$
shifts the phases of all Fourier components of $s$ by the angle
$\phi$.  By construction, $s^H_c$ (approximately) matches $s^L_c$ and
$h^L_c$ in time, phase and amplitude, where $s^L_c\doteq s^L_f$.

For the template-independent analysis of GW150914, $s^H_f$ and $s^L_f$
are directly compared to find $\Delta\phi$ and $A^{LH}$.  To do this,
construct from $s^H_f$ a complex signal
$S^H_f=s^H_f+i\,R(s^H_f,\pi/2)$.  Then cross-correlate $s^L_f$ and
$S^H_f$, and apply the same algorithm as for the template, above, to
determine the phase and amplitude that, when applied to $S^H_f$, will
yield $s^H_c$ that best matches $s^L_c\equiv s^L_f$.\vspace*{3pt}

The above steps yielded time offsets, amplitudes and phases that can
transform the original, unfiltered inputs $s^D$ and $H$ into new,
``coherent'' signals ${s'}^D$ and $H'$ that are approximately time and
phase coherent, and have approximately equal amplitudes, with
${s'}^L\equiv s^L$ as reference.  Iterative processing, using the
above steps but with ${s'}^D$ and $H'$ as input, provides significant
but rapidly diminishing corrections to the approximate results.

\begin{table*}[!h]
  \caption{\label{T:results}Parameters and results for each event.
    Definitions and calculation methods are given in
    Section~\ref{S:Extraction}.  The frequency pass bands were
    manually adjusted based on quality of fit.  The ``true signals''
    for simulated events have the event times, amplitudes and phases
    found for the actual events.  Uncertainties represent the
    $\pm 1$-sigma range of results from the simulated
    events.\vspace*{5pt}}
\vspace*{-8pt}
\rule{\textwidth}{.5pt}
\rule[10.5pt]{\textwidth}{.5pt}
\begin{minipage}{\textwidth}
\vspace*{-8pt}
  \begin{tabular*}{\textwidth}{l@{\extracolsep{\fill}}cccccc}
  Event & GW150914 & GW150914 & LVT151012 & GW151226 & GW170104\\
  \hline
  Template used & no & yes & yes & yes & yes\\
  Pass band~~[Hz] & 37 to 290 & 37 to 290 &
38 to 300 & 45 to 315 & 35 to 290\\
  Time diff.~(H-L)~~$\Delta t_{event}$ [ms] & $7.09^{+.20}_{-.09}$ & $7.11^{+.21}_{-.09}$ &
$-0.8^{+.6}_{-.6}$ & $0.6^{+.8}_{-.3}$ & $-3.1^{+.5}_{-.2}$\vspace*{2pt}\\
  Phase diff.~~$\phi^H-\phi^L$ [rad.] & $2.94^{+.22}_{-.27}$ & $2.95^{+.21}_{-.27}$ &
$2.7^{+.5}_{-.5}$ & $2.4^{+.5}_{-.4}$ & $3.2^{+.4}_{-.4}$\vspace*{2pt}\\
  H ampl.~~$ 10^3\,A^H$ & $1.30^{+.34}_{-.35}$ & $1.35^{+.30}_{-.44}$ &
$0.7^{+.3}_{-.5}$ & $1.4^{+0.6}_{-1.0}$ & $0.7^{+.2}_{-.3}$\vspace*{2pt}\\
  L ampl.~~$10^3\,A^L$ & $1.11^{+.29}_{-.27}$ & $1.11^{+.23}_{-.37}$ &
$0.7^{+.3}_{-.4}$ & $1.5^{+0.9}_{-1.0}$ & $0.6^{+.2}_{-.2}$\vspace*{2pt}\\
  Corr. coef.\footnote{For GW150914, the Pearson correlation
    between $s_w$'s calculated with and without use of the template is 0.996.}~~$r(s_w,h_{coh})$ & $0.985^{-.007}_{-.104}$ & $0.980^{-.006}_{-.113}$ &
$0.90^{-.05}_{-.30}$ & $0.86^{-.02}_{-.24}$ & $0.93^{-.01}_{-.19}$\vspace*{2pt}\\
  SNR~~$\rho_{ci}$ & $9.4^{-3.1}_{-5.1}$ & $7.1^{-1.4}_{-3.1}$ &
$1.9^{+0.5}_{-0.3}$ & $2.3^{-0.2}_{-0.7}$ & $3.5^{-0.5}_{-1.3}$\vspace*{2pt}\\
  SNR~~$\rho_{ti}$ & $9.1^{-1.9}_{-5.6}$ & $6.9^{-0.7}_{-4.0}$ &
$1.7^{+0.8}_{-1.1}$ & $1.9^{+0.7}_{-1.4}$ & $3.2^{+0.0}_{-1.8}$\vspace*{2pt}\\
  SNR~~$\rho^H_{tr}$ & $5.9^{-2.2}_{-4.3}$ & $5.3^{-1.8}_{-3.9}$ &
$1.9^{-0.4}_{-1.5}$ & $1.7^{-0.1}_{-1.3}$ & $2.0^{-0.3}_{-1.3}$\vspace*{2pt}\\
  SNR~~$\rho^L_{tr}$ & $4.0^{-1.1}_{-2.6}$ & $3.3^{-0.6}_{-2.2}$ &
$1.0^{+0.4}_{-0.6}$ & $1.0^{+0.4}_{-0.7}$ & $2.0^{+0.0}_{-1.2}$\vspace*{2pt}\\
  SNR~~$\rho_{tc}$ & $5.5^{-1.4}_{-3.8}$ & $4.9^{-1.2}_{-3.5}$ &
$2.0^{-0.3}_{-1.6}$ & $1.6^{+0.1}_{-1.2}$ & $2.5^{-0.3}_{-1.7}$\vspace*{2pt}\\
  SNR~~$\rho_{tw}$ & $5.6^{-1.5}_{-3.9}$ & $5.0^{-1.2}_{-3.6}$ &
$2.1^{-0.4}_{-1.7}$ & $1.7^{+0.0}_{-1.3}$ & $2.5^{-0.3}_{-1.7}$\\
\end{tabular*}
\end{minipage}
\rule{\textwidth}{.5pt}
\rule[10.5pt]{\textwidth}{.5pt}
\end{table*}

\subsection{Signal comparisons}

Comparing the waveforms $s^H_c$, $s^L_c$, $h^H_c$ and
$h^L_c$ reveals information about residual noise and allows
selection of combinations that best represent the gravitational waves.

The difference between $h^H_c$ and $h^L_c$ is not significant, so we
use their average:%
\beq%
h_{coh}=(h^L_c+h^H_c)/2\,.%
\eeq%
Subtracting the template from the signals gives the residuals:%
\beq%
r^D_c=s^D_c-h_{coh}\,.\label{Eq:resid}%
\eeq%
We define the (time-domain) template/residual SNR for detector $D$ by
\beq%
\rho^D_{tr}=\sigma(h_{coh})/\sigma(r^D_c)\,,\label{Eq:rhoD}%
\eeq%
where $\sigma(\cdot)$ denotes rms over the time interval spanned by
all the \emph{t-f} bands.\footnote{More explicitly, we define the mean
  $X$ and variance $\sigma^2$ of a time series $X_i$ with $N$ samples
  as $X=\frac{1}{N}\sum_{i=1}^N X_i$ and
  $\sigma^2(X_i) = \frac{1}{N}\sum_{i=1}^N (X_i-X)^2$. Here the
  samples are summed over the time interval spanned by all of the
  \emph{t-f} bands, and the mean values are typically close enough to
  zero that $X$ effectively vanishes.}  A combined signal, that
weights each $s^D_c$ inversely with the noise indicated by
$\sigma(s^D_{cbp})$, is:%
\beq%
s_w=\frac{s^H_c+ r\,s^L_c}{1+r}\,,%
\eeq%
where $r=\sigma(s^H_{cbp})/\sigma(s^L_{cbp})$.  Equally-weighted
coherent and incoherent combinations of the detector signals are:%
\beq%
s_{coh}=(s^L_c+s^H_c)/2\,,\label{Eq:s_coh}%
\eeq%
\beq%
s_{inc}=(s^L_c-s^H_c)/2\,.\label{Eq:s_inc}%
\eeq%
For detectors with uncorrelated noise and similar true signals $g$,
$s_{inc}$ should be noise-dominated.  It is reasonable to expect noise
at a similar level to $s_{inc}$ to be hidden in $s_{coh}$ and $s_w$,
and in the coherent and weighted residuals:%
\beq%
r_{coh}=s_{coh}-h_{coh}\,,\label{Eq:r_coh}%
\eeq%
\beq%
r_w=s_w-h_{coh}\,.\label{Eq:r_w}%
\eeq%
Thus, in the SNR values defined below, $\sigma(s_{inc})$ should be
interpreted as a proxy for the rms value of the hidden coherent noise.

New SNR definitions follow from the above:%
\beq%
\rho_{tw}=\sigma(h_{coh})/\sigma(r_w)\,,\label{Eq:rhotw}%
\eeq%
\beq%
\rho_{ti}=\sigma(h_{coh})/\sigma(s_{inc})\,,\label{Eq:rhoti}%
\eeq%
\beq%
\rho_{tc}=\sigma(h_{coh})/\sigma(r_{coh})\,,%
\eeq%
\beq%
\rho_{ci}=\sigma(s_{coh})/\sigma(s_{inc})\,.\label{Eq:rhoci}%
\eeq%

The above SNR values measure, in slightly different forms, the
strength of the extracted signals or scaled templates relative to
detector noise or computed residuals. They are mathematically and
conceptually distinct from the matched-filter \emph{detection} SNRs,
reported in the discovery papers~\cite{PhysRevLett.116.061102,
  PhysRevD.93.122003, PhysRevLett.116.241103, PhysRevX.6.041015,
  PhysRevLett.118.221101,PhysRevLett.119.141101,2017ApJ...851L..35A}.
Section \ref{Ss:SNR} has further remarks regarding these different
ways of measuring SNR.

For a measure of the overlap between the extracted signal and template
we compute the Pearson correlation coefficient:%
\beq%
r(s_w,h_{coh})=\frac{s_w\cdot h_{coh}}{\sqrt{(s_w\cdot
    s_w)(h_{coh}\cdot h_{coh})}}.\label{Eq:Pearson_r}%
\eeq%

Before presenting results, we will describe the use of simulations to
validate and statistically characterize our signal extraction
methodology.

\subsection{Simulations}\label{SS:Simulations}

For each analyzed event, inserting the best fit amplitude $A^D$ and
phase $\phi^D$ into Eq.~(\ref{Eq:template}) yields the real-valued
template $h^D$ for the unfiltered gravitational wave signal at site
$D$.  Simulated event records $q^D$ were created by adding $h^D$ to
each of 252 overlapping 32~s noise records drawn from the 4096~s
record $s^D_l$ (avoiding the time of the actual event).  Each pair
$(q^H,\,q^L)$ was then processed in exactly the same way as the
records containing the real event.  However, instead of being
individually tuned, all the simulated events used the same pass band
and the same width parameter for \emph{t-f} bands as were used for the
corresponding real event.

The weighted waveforms $q_{w,i}$, where $i$ indicates the simulation
number, extracted from the simulated events can be directly compared
to the known injected signal $h_{coh}$.  Statistical distributions of
parameters determined in the signal extraction process\,---\,time
offsets, $A^D$, $\phi^D$, and SNR values\,---\, are indicative of the
uncertainties of the parameters found for the real events.

The signal extraction process failed to find an event time offset
between detectors within the acceptable $\pm 10$~ms range for 14 of
the 1260 simulated events, one failure for each of GW150914 (template
not used) and GW151226 and the rest for LVT151012.  These were omitted
from further analysis.  Boosting the injected signal amplitudes by
20\% reduced the number of failures to four, by 30\% to one, and by
80\% to zero.

\begin{figure*}[!t]
  \center\includegraphics[height=2.5in]{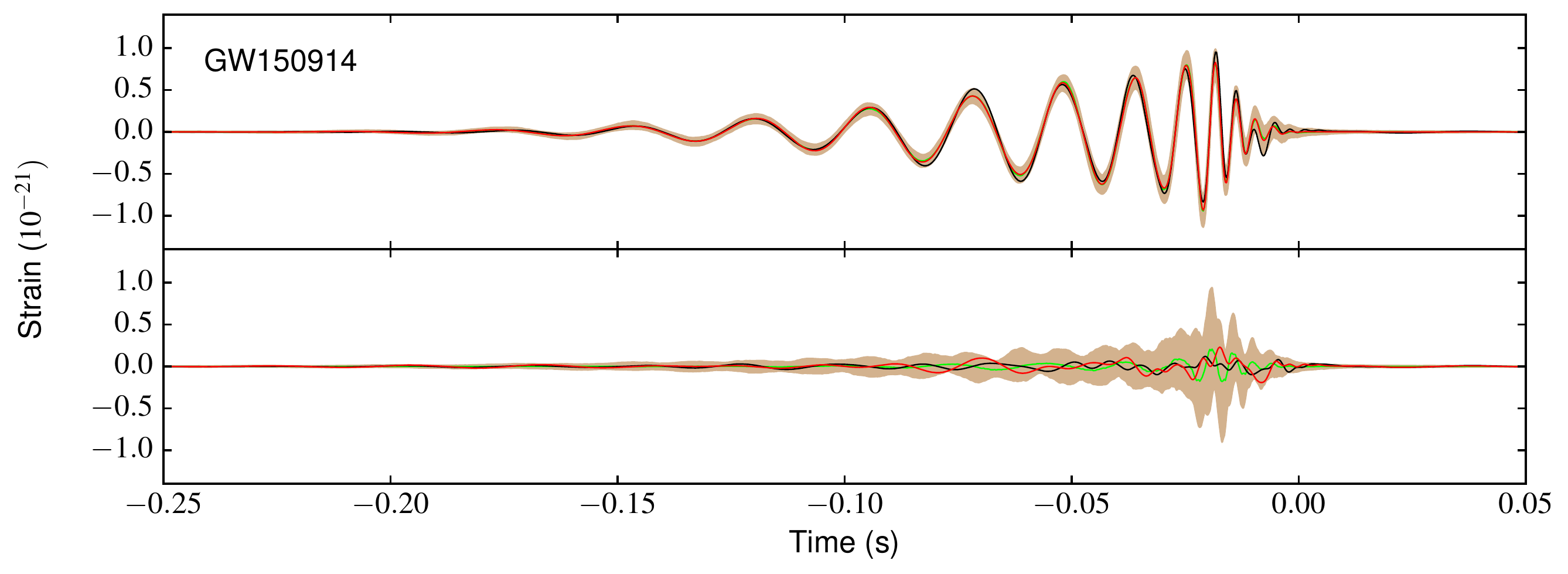}
  \center\includegraphics[height=2.5in]{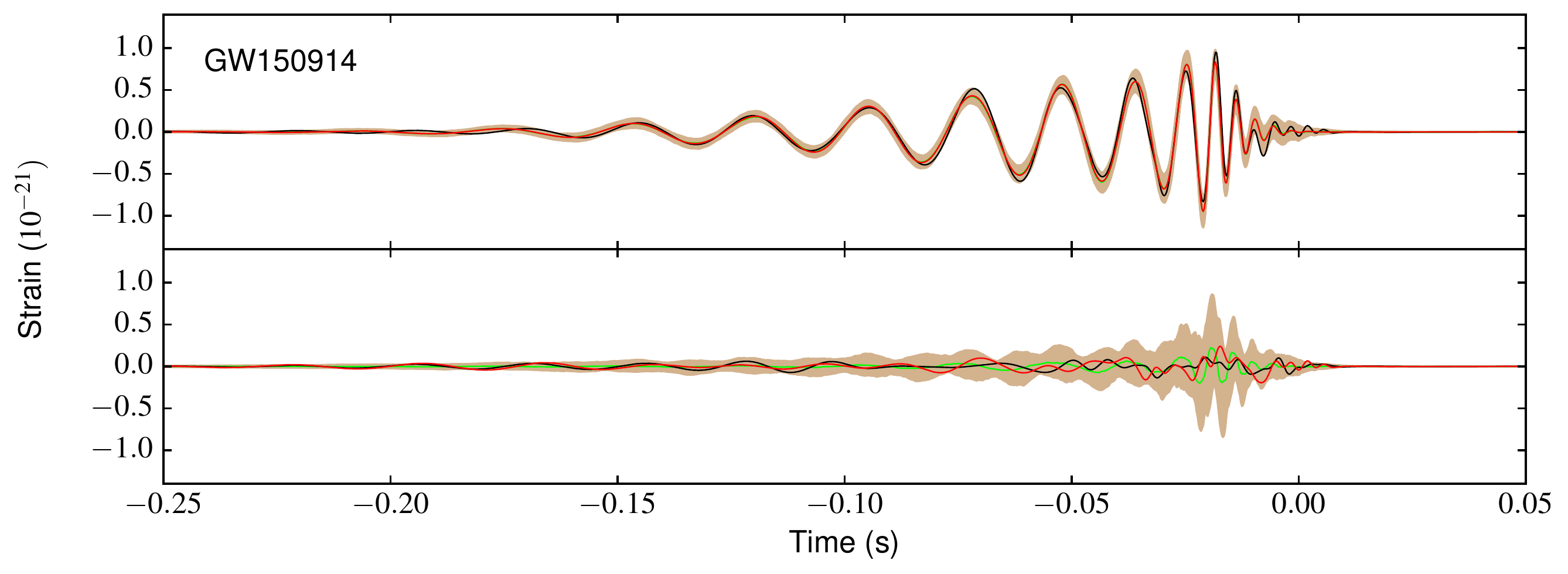}
  \caption{GW150914 without (upper) and with (lower) use of the
    template. Top panels: weighted combined signal $s_w$ (black) and
    coherently matched template $h_{coh}$ (red), with Livingston as
    reference.  When the template was added to noise records, 90\% of
    the extracted signals $q_{w,i}$ lie within the light brown band.  The
    median of the $q_{w,i}$ signals from the simulations is a green line
    almost entirely hidden by the red line.  Bottom panels: incoherent
    signal $s_{inc}=\scriptstyle\frac{1}{2}\textstyle(s^L_f-s^H_f)$
    (black) and residual $r_w=s_w-h_{coh}$ (red).  Equivalent
    residuals for simulated events lie within the light brown band
    90\% of the time; the green line is the median simulation
    residual.}
  \label{F:GW150914_TemplateStats}
\end{figure*}

\subsection{Results}\label{SS:Results}

Values of the parameters, correlation coefficients
(Eq. (\ref{Eq:Pearson_r})), and SNRs (Eqs.~(\ref{Eq:rhoD}),
(\ref{Eq:rhotw})\,--\,(\ref{Eq:rhoci})) found for each of the events,
with corresponding $\pm 1$-sigma bounds derived from the simulations,
are given in Table \ref{T:results}.  For the time differences, phase
differences, and amplitudes, results for the actual events\,---\,which
were used as ``true'' parameter values for the simulations\,---\,lie
reasonably within the statistical ranges found from simulated events.
For the correlation coefficients and SNR values, results from actual
events generally lie above the 1-sigma ranges from
simulations\,---\,we explain why in Section \ref{S:Discussion}.

\begin{figure*}[!t]
  \center\includegraphics[height=2.5in]{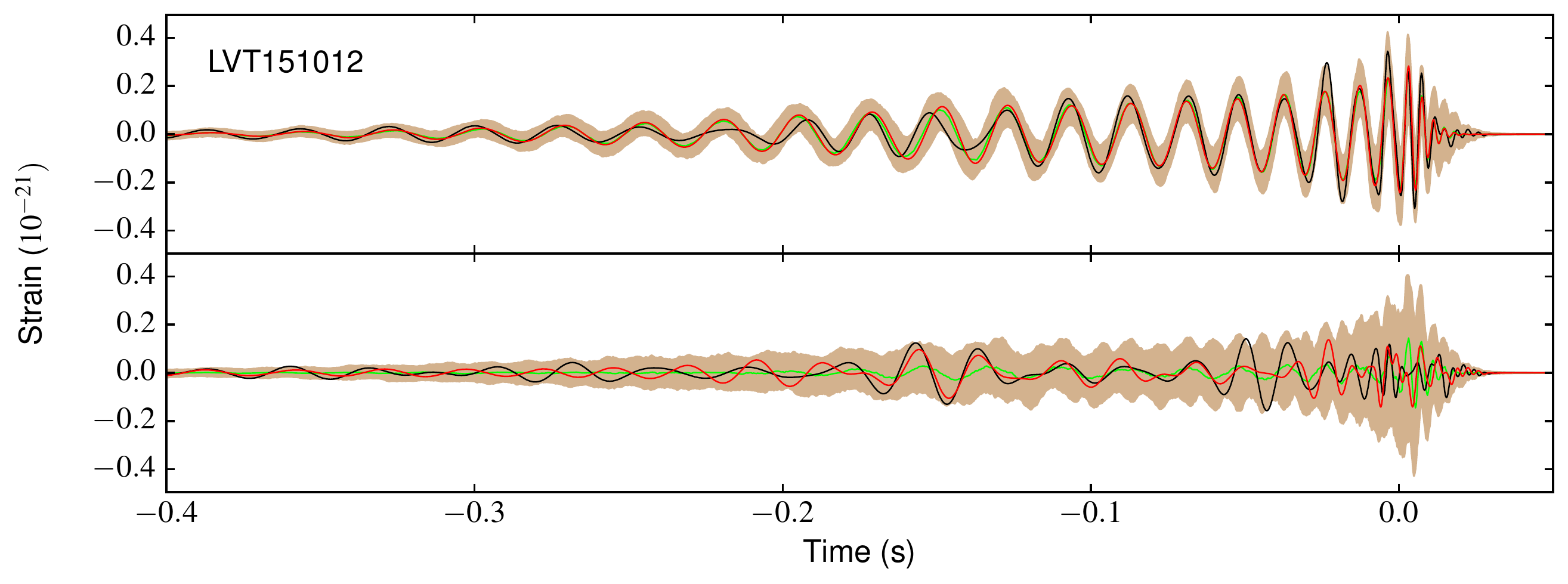}
  \center\includegraphics[height=2.5in]{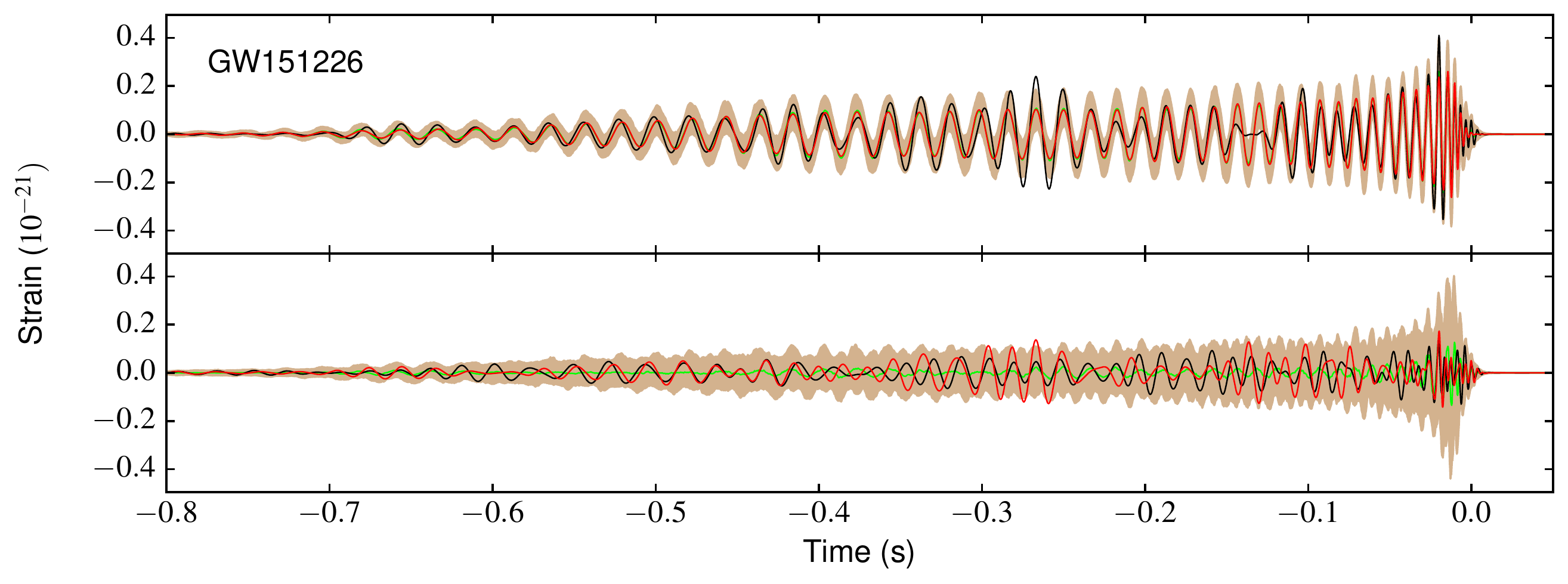}
  \center\includegraphics[height=2.5in]{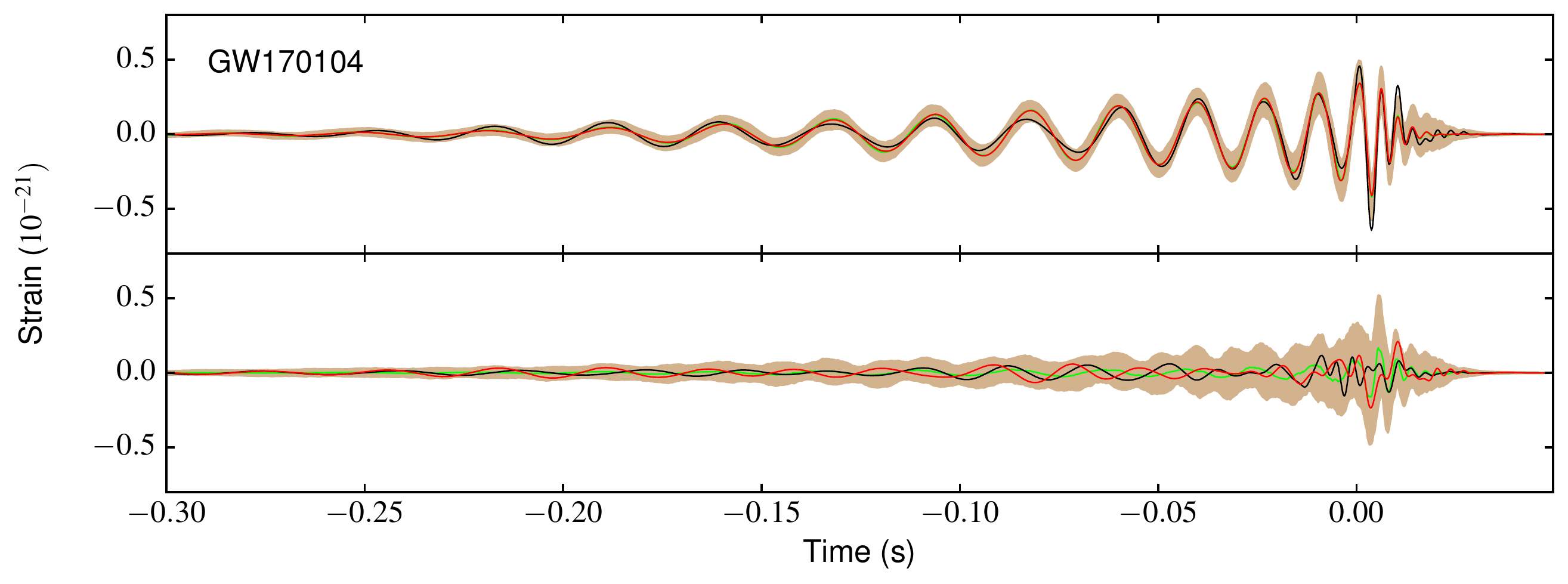}
  \caption{LVT151012, GW151226 and GW170104, all with use of
    templates.  See legend of Fig. \ref{F:GW150914_TemplateStats} for
    details.}
  \label{F:TemplateStats}
\end{figure*}

For GW150914, analyzed both without and with use of the template,
extracted waveforms, and bands enclosing 90\% of the extracted
waveforms from simulations, are shown in
Fig.~\ref{F:GW150914_TemplateStats}.  Similar plots for LVT151012,
GW151226 and GW170104 are shown in Fig.~\ref{F:TemplateStats}.  Time
periods when $r_w$ and $s_{inc}$ are nearly equal or nearly opposite
will have the noise at Hanford or Livingston dominating over the
other.  Event residuals lie almost completely within the 90\% bands
obtained from the simulations.

Plotted as green lines in the upper panels, the median values of the
extracted waveforms $q_{w,i}$ from the simulations are almost entirely
obscured behind red lines for $h_{coh}$.  In the same order as the
plots of Fig.~\ref{F:GW150914_TemplateStats} and
\ref{F:TemplateStats}, the correlation coefficients between $h_{coh}$
and the simulation medians are 0.9983, 0.9997, 0.9918, 0.9968, and
0.9987.  This indicates that the extraction process gives an unbiased
(but still noisy) representation of the true gravitational wave
signal.

The waveforms $s_w$ obtained for GW150914 without and with use of the
template are nearly identical, with 0.996 correlation.  For all
events, the extent of the light brown band in the upper panels seems
consistent with the $\pm$ variation of time differences, phase
differences, and amplitudes (Table \ref{T:results}).  The simulation
residuals shown by the light brown bands in the lower panels have high
amplitudes near the zero crossings of $s_w$, suggesting (known) phase
errors as the main cause.

\section{Discussion}\label{S:Discussion}

Matched filters are effective at discovering the occurrence of an
\emph{expected} signal $h$ within a noisy data stream $s$.  If
$s=g+n$, where the \emph{true} signal $g$ correlates strongly with
$h$, then the matched filter will indicate detection of $h$.  However,
the matched filter does not distinguish the difference $\delta=g-h$
from the noise.

In this work, the existence, times and basic characteristics of the
reported gravitational wave signals have been taken as
given\,---\,prior knowledge.  Working with data provided by the LOSC,
we have developed a filtering strategy to preferentially reduce $n$
but not $g$, thereby making the filtered signal $s_f$ a better
representation of $g$.  This was done without imposing specific phase
relationships between the different spectral components.  However, our
assumption that the signal has the characteristic form expected of
black hole coalescence\,---\,as seen in the templates\,---\,imposed
the implicit requirement that the phases vary smoothly with frequency.

Our filtering strategy is grounded in the recognition that the true
signal $g$ in each narrow frequency band $\Delta f_i$ of a black hole
coalescence event has significant magnitude only within a brief,
contiguous time interval $\Delta t_i$.  Measured signal $s$ in the
band $\Delta f_i$ but outside the interval $\Delta t_i$ will be
dominated by noise.  The algorithms we have developed determine
suitable \emph{t-f} bands directly from the observed, sufficiently
strong signal (GW150914) or from a reasonable-fit template provided by
LIGO.  The bands overlap in both frequency and time to ensure almost
all of the true signal is maintained in the filtered result.
Normalization compensates for band overlaps, ensuring that no spectral
component of the output has weight greater than one.  By adjusting the
width of the time windows, using parameter $\alpha$, noise rejection
can be balanced against distortion of the true signal (and template).
Choosing large $\alpha$, for strong signals, gives time windows that
include almost all true signal energy.  Choosing a smaller $\alpha$,
for weak signals, accepts some filtering of true signal in order to
increase noise suppression.  However, Fig.~\ref{F:timeFreq2} shows
that even the smallest value, $\alpha=1.215$, preserves most of the
signal.

Correlation coefficients $r(q_{w,i},h_{coh})$ for the simulated events
were generally smaller than $r(s_w,h_{coh})$ for actual events.  To
explain this, we note that the templates $H$, provided by LIGO, are
particularly good matches for the observed strains $s^D=n^D+g^D$,
which differ from the true signals $g^D$.  If the $g^D$ could be
known, then applying the LIGO analysis process to simulated signals
$q^D_i=n^D_i+g^D$ would lead to many different templates $H_i$, each
tailored to the unique noise $n^D_i$.  But in our simulations the true
$g^D$ are not known.  As a proxy, we have used instead $h^D$, derived
from $H$ and thus tainted by $n^D$ at the time of the actual event.
The difference $g^D-h^D$ is not ignorable.  The result is that
waveforms $q_{w,i}$ extracted from the simulated signals tend to not
match $h_{coh}$ as well as $s_w$ does.

Since we have no evidence that the (unknown) noise $n^D$ at the time
of the actual event is statistically different from the noise samples
$n^D_i$, the distribution of extracted waveforms $q_{w,i}$ obtained
from the simulations should be taken as a good indication of the
uncertainty with which $h_{coh}$ (and $s_w$) represent $g$.  Because,
for simulation of a given event, the same signal $h^D$ was injected
into each noise sample $n^D_i$, the median of the $q_{w,i}$ should
closely match $h_{coh}$, as seen in Figures
\ref{F:GW150914_TemplateStats} and \ref{F:TemplateStats}.  For the
same reason, the distributions of simulation results for the time
offset, phase and amplitude parameters should have mean values that
closely approximate the corresponding parameters for the actual
events, from which the $h^D$ were constructed.  The noise that gives
these distributions their widths also makes uncertain the true values
of the parameters, and hence the means of the distributions that would
result if $h_{coh}$ corresponded precisely to $g$.  For each given
parameter, the uncertainty of the mean stems from the same noise, and
will thus have the same distribution, as the parameter values obtained
from simulations.  This implies that uncertainties of the true
parameter values have bands $\sqrt{2}$ wider than indicated by the
$\pm 1$-sigma ranges from the simulations.

\subsection{Remarks on SNR}\label{Ss:SNR}

\newlength{\digitwidth} \settowidth{\digitwidth}{0} We have defined
SNRs (Eqs.~(\ref{Eq:rhoD}), (\ref{Eq:rhotw})--(\ref{Eq:rhoci})) that
are ratios of rms amplitudes of filtered time series records that
serve as proxies for the true signal and true noise.  Computed in the
time domain, these are sometimes referred to as \emph{instantaneous}
SNRs.  Results for the studied events are shown in
Table~\ref{T:results}.

For GW150914, slightly higher SNR values for the signal extraction
without, as opposed to with, use of the template can be attributed to
the somewhat greater noise exclusion by narrower time windows for the
\emph{t-f} bands, evident in Fig.~\ref{F:timeFreq}.  This difference
is also seen in the correlation coefficients, but there is no
significant difference in the obtained signal parameters.

For all events but LVT151012, the values of $\rho_{ci}$ and
$\rho_{ti}$, with $\sigma(s_{inc})$ in the denominator, are higher
than the other SNR measures, which use residuals as proxies for noise.
This can be traced to contributions $g-h$ to the residuals, that
become relatively more important when the detector noise (for which
$s_{inc}$ is a proxy) is small.

More notably, SNR values for simulated events tend to be lower than
for actual events, just as with correlation coefficients.  Again, the
main explanation is that the template $H$, provided by LIGO, is a
statistically better match for the observed strains $s^D=g^D+n^D$ than
it is for the (unknown) true strains $g^D$, thus giving anomalously
high SNRs.  For the simulated signals, the injected signal $h^D$ is
derived directly from $H$ and there is nothing to make the template
match $q^D_i= h^D+n^D_i$ better than it does $h^D$.  The SNRs from
simulations should thus be considered a more reliable representation
of the truth.

We further observe that although $n^D$ makes a tainting contribution
to $s_w$ and $H$, this will not significantly diminish the coherent
component of noise that is complementary to $s_{inc}$ (or $q_{inc,i}$)
and which we have assumed has rms value similar to $\sigma(s_{inc})$.

The instantaneous SNRs discussed above are distinct in definition,
purpose and interpretation from a matched-filter SNR.  The
matched-filter SNR $\rho$, used by LIGO to quantify the match between
a test template $(\mathfrak{h}_p,\mathfrak{h}_c)$ and a noisy measured
signal $s$, for a given detector, is defined
by~\cite{PhysRevD.49.2658, PhysRevX.6.041015}:%
\beq%
\rho^2(t)=[\langle s|\mathfrak{h}_p\rangle^2(t)+\langle
s|\mathfrak{h}_c\rangle^2(t)]\,.\label{Eq:rho2}%
\eeq%
Here, the optimally normalized correlation between $s$ and each
template component, $\mathfrak{h}_p$,\,$\mathfrak{h}_c$, offset in time
by $t$, is defined by:%
\beq%
\langle s|\mathfrak{h}\rangle(t)= 4\,\Upsilon\,
\mathrm{Re}\int_0^\infty\frac{\tilde{s}(f)\tilde{\mathfrak{h}}^*(f)}
{S_n(f)}e^{2\pi i f t}df\,,\label{Eq:corr}%
\eeq%
where $S_n(f)$ is the positive frequency PSD of the detector noise,
and $\Upsilon$ is a normalization factor.  For each detector,
$\Upsilon$ is chosen such that for an ensemble consisting of many
distinct realisations of the detector noise $s$ (with no signal),
taken from the time interval used to determine $S_n(f)$, the ensemble
averages of
$\langle s|\mathfrak{h}_p\rangle^2=\langle s|\mathfrak{h}_c\rangle^2 =
1$, independent of $t$.  With no signal present, Eq.~(\ref{Eq:rho2})
then gives the (time-independent) expected value
$<\!\rho^2\!>\,\,=2$.\footnote{See~\cite{PhysRevD.84.122004} for
  discussion of matched filters in the context of discrete time series
  and non-Gaussian noise.}

A gravitational wave event adds real signals $g$ to the noise at
each of the detectors, resulting in peaks $(\rho^2)^H$, $(\rho^2)^L$
of $\rho^2(t)$ at times $t^H\!,\,t^L$ separated by not more than the
light travel time between detectors.  Combining the peak $\rho$ values
from the two detectors in quadrature gives the event SNR:\\
\beq \rho_{ev}=\sqrt{(\rho^2)^H + (\rho^2)^L}\,.
\label{Eq:rhoEvent}
\eeq

The $\rho_{ev}$ values published by LIGO for the four events studied
here are~\cite{PhysRevX.6.041015,PhysRevLett.118.221101}:\\[-6pt]
\begin{center}
  \begin{tabular}{lcr}
    GW150914&--&23.7\\
    LVT151012&--&9.7\\
    GW151226&--&13.0\\
    GW170104&--&13.\hspace*{\digitwidth}
  \end{tabular}
\end{center}
These values represent the collective result of analysis with many
different templates, each corresponding to different physical
parameter values $\theta$, and the requirement for detection of
similar signals at both detectors with sufficiently small time offset.
As described in \cite{PhysRevD.49.2658}, maximum-likelihood parameter
values $\theta_{ML}$ and uncertainties have been estimated by
comparing Bayesian evidence for a wide range of plausible parameter
values.  The template $h[\theta_{ML}]$ corresponding to $\theta_{ML}$
then maximizes the event SNR.  For GW150914, instead of the 23.7 cited
above, consideration of a finer sample space for $\theta$ gave
$\rho_{ev}[\theta_{ML}]=25.1^{+1.7}_{-1.7}$
\cite{PhysRevLett.116.241102}.

The precise values of matched-filter SNRs obtained using
Eqs.~(\ref{Eq:rho2}) and (\ref{Eq:corr}), and the likelihoods obtained
in the Bayesian analysis, will always depend on the \emph{actual}
noise $n_a$ realized during the time interval $(t_0-\Delta t,t_0)$
when the signal $g$ has significant amplitude.  Different noise will
yield different SNRs.  To demonstrate this, we have computed
$\rho_{ev}$, as defined by Eq.~(\ref{Eq:rhoEvent}), for each of the
studied events.  Using the LIGO-provided templates
$(\mathfrak{h}_p,\mathfrak{h}_c)$, we found slightly lower $\rho_{ev}$
values than those listed above.  For each of the events, we then
injected the same signal $h$ (our best fit) into many different noise
realizations $n_i$ (as in our simulations), and found that
matched-filter SNRs $\rho_{ev,i}$ for the template
$(\mathfrak{h}_p,\mathfrak{h}_c)$ have variance
$\mathrm{var}(\rho_{ev,i})\sim 1$\,.  The results of our calculations
are:\\[-6pt]
\begin{center}
  \begin{tabular}{lccc}
    Event&$\rho_{ev}$&$\mathrm{mean}(\rho_{ev,i})$&$\mathrm{var}(\rho_{ev,i})$\\
    \hline\\[-10pt]
    GW150914&23.5&24.0&1.04\\
    LVT151012&\hspace*{\digitwidth}8.6&\hspace*{\digitwidth}8.2&0.83\\
    GW151226&11.2&10.9&1.06\\
    GW170104&12.6&11.8&1.07
  \end{tabular}
\end{center}
SNRs for simulated events at the individual detectors have similar
variances:
$\mathrm{var}(\sqrt{(\rho^2)_i^H}) \sim
\mathrm{var}(\sqrt{(\rho^2)_i^L}) \sim 1$.  Injecting the same $h$
into noise records that are identical except for relative time shifts
as small as 2~ms gives detection SNRs whose variations about the mean
are apparently uncorrelated.

For the studied events, the actual data $s=g+n_a$ have limited useful
bandwidth; and the signal $g$ is detectable only for a brief interval
of $\sim\!\!1$~s or less, which is too short to demonstrate the
(band-limited) Gaussian nature of the noise.  With the limited
$1/\Delta t$ frequency resolution afforded by such a brief interval,
the fluctuations of $|\tilde{n}_a|$ about $\sqrt{S_n}$ cannot exhibit
the obvious stochastic character exemplified by $\tilde{s}_{cbp}$ (a
32~s record) in the left panel of Figure \ref{F:FFT}.

\begin{figure}[!t]
  \center\includegraphics[height=2.5in]{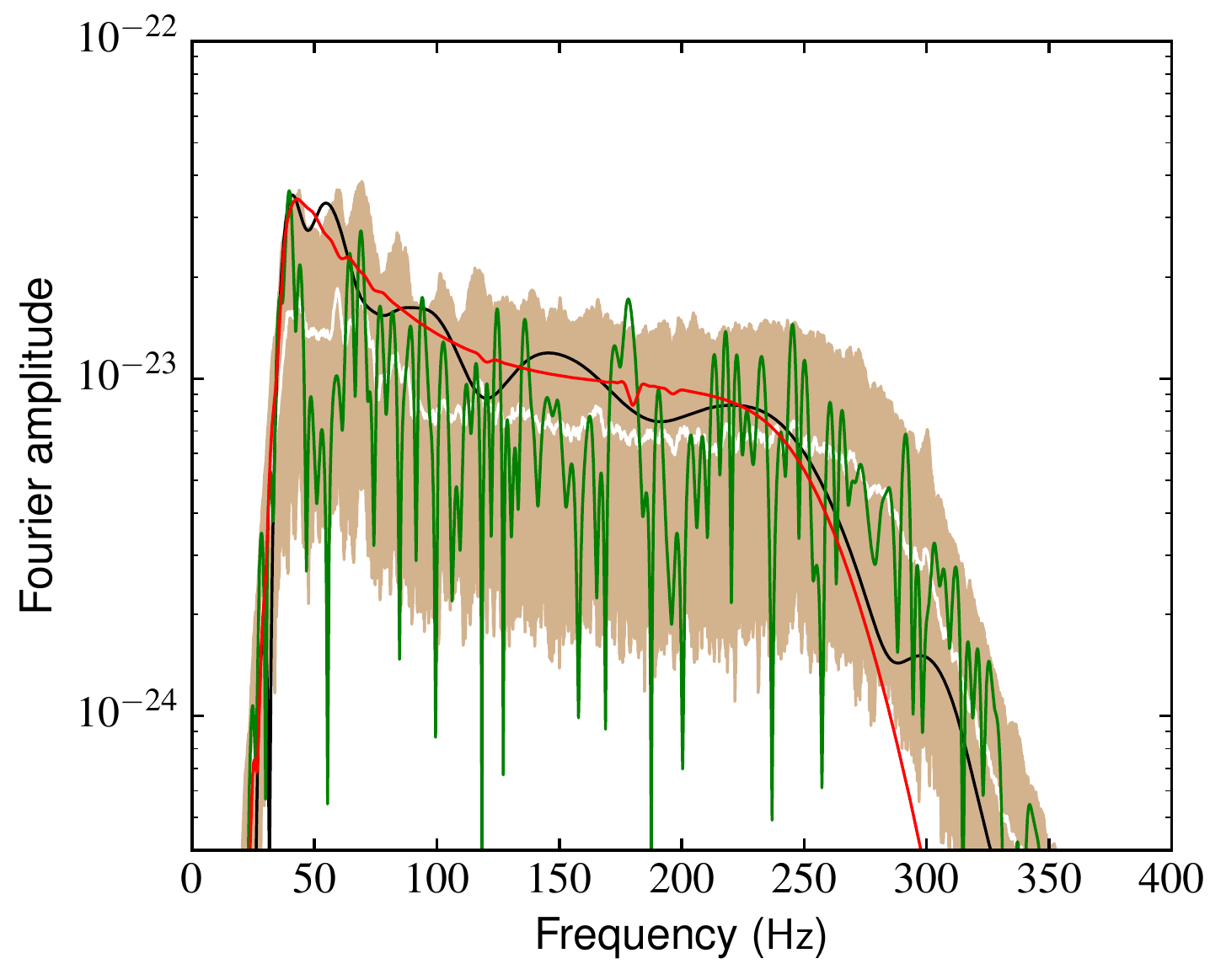}
  \caption{GW150914 Fourier amplitudes.  The white line shows the
    median of $|\tilde{n}_i|$ for 252 noise samples $n_i$ from
    Livingston; 90\% of the $|\tilde{n}_i|$ lie within the light brown
    band.  FFTs were computed using 32~s records, with a smoothly
    tapered window $W$ masking all but 0.4~s.  Amplitudes are scaled
    by $\sqrt{32/0.3}$ to compensate for $W$ and allow comparison with
    $\sqrt{S}$.  The smooth red line shows the template spectrum,
    $|\tilde{h}_{cbp}|$, with the same $W$ and scaling.  The black
    line shows $|\tilde{s}_w|$ and the jagged green line
    $|\tilde{n}|$, both also scaled, where $n=(s^L_{cbp}-s_w)\times W$
    is the noise filtered out using the \emph{t-f} bands to obtain
    $s_w$.}
  \label{F:GW150914_FourierAmpl}
\end{figure}

To illustrate this, Fig.~\ref{F:GW150914_FourierAmpl} shows Fourier
amplitudes for a 0.4~s period spanning the GW150914
event.\footnote{Window tapers reduce the effective sample length to
  0.3~s.}  The spectrum of the extracted signal $s_w$ matches that of
the template $h_{cbp}$ (for the same 0.4~s) quite well; differences
between them can be attributed to both real differences $\delta=g-h$
and residual coherent noise.  The noise in $s_{cbp}$, i.e., what
remains after subtracting the signal $s_w$, has spectral fluctuations
consistent with the 90\% band of $|\tilde{n}_i|$ obtained by analyzing
252 different 0.4~s noise samples $n_i$.  The median of the
$|\tilde{n}_i|$ has much smaller fluctuations, making it similar
(except for the imposed pass band) to $\sqrt{S_b(f)}$ in
Fig.~\ref{F:FFT}.

The spectra in Fig.~\ref{F:GW150914_FourierAmpl} have been normalized
to achieve a frequency-domain signal power (total energy / 32~s) that
matches the mean time-domain power (corrected for window tapers)
during the chosen 0.4~s segment of each 32~s record.  Every 0.4~s
segment of stationary noise should have similar power\,---\,so it is
not surprising that the median amplitude of $\sim\!10^{-23}$ matches
$\sqrt{S_b(f)}$ quite well.  The correspondence for $|\tilde{s}_w|$
and $|\tilde{h}_{cbp}|$ is more subtle.  The 0.4~s window was chosen
to snugly frame the event.  Doubling the window length would have cut
the signal and template Fourier amplitudes in half; reducing the
length would have excluded portions of the signal and changed the
shape of its spectrum.  Changing the window length would also have
sampled different noise and given a different detailed noise spectrum.

We expect that the above-demonstrated dependence of matched-filter SNR
values on the actual detector noise coincident with a black hole
coalescence event will apply also to the outcomes of Bayesian
parameter estimation.  Similar to our use of simulations for
estimation of uncertainties, the frequentist approach of injecting the
same signal into different noise samples might provide useful guidance
on the noise-related uncertainties of estimated physical parameters.
At the very least, it would test the assumption that Bayesian and
frequentist approaches give similar outcomes.

For the purposes of the present work, matched-filter SNR values
indicate the trust that can be placed in the prior knowledge our
signal extraction method relies on.  Our instantaneous SNR values then
indicate how strongly the extracted signal can be made to stand above
the residual noise, when the prior knowledge is used to guide the
definition of \emph{t-f} bands for selective filtering.  The
matched-filter and instantaneous SNR values are thus seen as
complementary.

\subsection{Comparison with wavelet approaches}\label{Ss:Wavelet}

Our analysis method, as presently implemented, is limited to the
narrow purpose of extracting clean representations of gravitational
wave signals from already identified black hole coalescence events.
Although, when the signal is weak, we make use of reasonable-fit
templates, the prior knowledge we use from the template is only the
approximate evolution of frequency over time\,---\,to allow selection
of \emph{t-f} bands.

Wavelet based approaches~\cite{0264-9381-32-13-135012,
  1742-6596-363-1-012032} offer flexibility in analyzing a wide
variety of gravitational wave signals.  Such methods can be used for
both signal detection and characterization.  Combining wavelet and
template based methods can improve parameter estimation and rejection
of instrument glitches.

For two events we can give quantitative comparisons.  For GW170104,
the wavelet based analysis achieved 87\% overlap between the
maximum-likelihood waveform of the binary black hole model and the
median waveform of the morphology-inde-pendent
analysis~\cite{PhysRevLett.118.221101}.  We found 93\% overlap between
the LIGO-provided template and the extracted signal $s_w$, and 99.87\%
overlap between the template injected into real noise records and the
median waveform extracted from the 252 simulated events.  For
GW150914, processed without use of the template, our corresponding
overlaps are 98.5\% and 99.83\%, while wavelet analysis achieved a
94\% overlap with the binary black hole model
\cite{PhysRevLett.116.241102}.  However, as we have noted above, the
overlaps with the true gravitational wave signals $g$ are almost
certainly lower because the template (or maximum-likelihood) waveforms
are tainted by the actual noise at the time of detection.

\subsection{Correlation of noise and residuals}

It has been argued in~\cite{2017JCAP...08..013C} that
cross-correlations of the residuals at the two detectors have peaks
corresponding to the event time offsets for GW150914, GW151226 and
GW170104.  This was cited as support for the claim ``A clear
distinction between signal and noise therefore remains to be
established in order to determine the contribution of gravitational
waves to the detected signals.''  For GW150914, correlations due to
the calibration lines in the vicinity of 35~Hz were cited as further
support.

We have argued in Section \ref{S:Noise} that apparent failure to use
appropriate window functions casts doubt on the conclusions
of~\cite{2017JCAP...08..013C}.  Our analysis indicates, nonetheless,
that cross-correlations of residuals may indeed have significant peaks
corresponding to the event time offsets\,---\,but the origin of these
peaks is not correlations between noise at the two detectors.
Instead, they arise because the true signal is not identical to the
template, and the difference $\delta=g-h$ will be included in both the
residuals.  Ignoring minor differences between the true signals, and
between corresponding templates, at the two sites (after adjusting the
Hanford time, phase and amplitude), we can expand the individual site
residuals as $r^D=n^D+\delta$.  The weighted residuals
(Eq.~(\ref{Eq:r_w})) then become $r_w=w\,n^L_c+(1-w)n^H_c+\delta_c$,
where the subscript $c$ denotes the same processing as used to obtain
$s^D_c$ and the weight $w\sim 0.5$.  Since the noise signals $n^H$ and
$n^L$ are independent (and thus randomly correlated), their
antisymmetric combination $n_{inc}\doteq s_{inc}=(n^L_c-n^H_c)/2$ will
have similar statistics to, but be independent of, $n_w=r_w-\delta_c$.
It follows that, if $\delta_c$ is significant,
$\sigma(r_w)>\sigma(s_{inc})$.  From the lower panels of Figures
\ref{F:GW150914_TemplateStats} and \ref{F:TemplateStats} it appears
that, for the actual events, $\sigma(r_w)\simeq \sigma(s_{inc})$.
However, the light brown bands show that the simulation residuals can
have significantly greater magnitude.  We have attributed this to
parameter errors (especially phase) in the extracted $q_{w,i}$'s, with
attendant contributions to $\delta_c$.  The common contribution of
$\delta_c$ to $r^H_c$ and $r^L_c$ will inevitably produce a peak at
the event time offset in the cross-correlation of residuals, even
though $n^H_c$ and $n^L_c$ may be randomly correlated.

\section{Conclusions}

We have introduced a method for extracting, from noisy strain records,
clean gravitational wave signals for black hole coalescence events.
Our method takes as prior knowledge the existence and timing of the
events, established through matched-filter and other techniques by
LIGO.  For events whose signals cannot be readily detected without
matched-filter techniques we also use, as prior knowledge, the
approximate time-dependent frequency content of a reasonable-fit
template waveform provided by LIGO.

Statistical tests show that, after filtering to remove spectral peaks
and restrict the pass band, the recorded signals prior to and
following the detected events are similar to band-passed Gaussian
noise.  There is no evidence of correlation between the two detectors
that would indicate a causal connection.

The extracted waveforms, including for GW150914 extracted without
using the template, correlate very strongly with the LIGO templates.
Simulations, with template-derived signals injected into real noise
records and processed the same as the actual events, show that the
analysis method is unbiased.  But the simulation results give weaker
correlation with the template and lower SNR values than the real
events.  This suggests that the template is tainted by noise at the
time of the real event\,---\,it matches the real signal plus noise
$(g+n)$ better than it does $g$ alone.  The simulation results, based on
many different stretches of real noise, support more realistic
estimates of parameter uncertainties and SNR values for the actual
events.

The methods and results presented here are complementary to the
matched-filter, wavelet and statistical analyses used by LIGO.
Combining these approaches may lead to more robust determination of
the physical parameters and uncertainties for black hole coalescence
events.  Generalizing our methods to different kinds of physical
events, and even to very different problems, may also prove fruitful.

\section*{Acknowledgments}

The authors thank Louis Lehner and Sohrab Rahvar for helpful
discussions, and the three anonymous referees whose feedback has led
to significant improvements over the original draft.  This research
has made use of data, and software to load it, obtained from the LIGO
Open Science Center (https://losc.ligo.org), a service of LIGO
Laboratory and the LIGO Scientific Collaboration. LIGO is funded by
the U.S. National Science Foundation.  Research at the Perimeter
Institute for Theoretical Physics is supported by the Government of
Canada through Industry Canada and by the Province of Ontario through
the Ministry of Research and Innovation (MRI).

\section*{References}

\bibliography{GWrefs}

\end{document}